\documentclass[a4paper]{jpconf}

\usepackage{graphicx}
\usepackage{dcolumn}

\usepackage{bm} 
\usepackage{amsmath}
\usepackage{amssymb}
\usepackage{hyperref}
\usepackage{tensor}

\newcommand{\pa}{\partial}
\newcommand{\ba}{\begin{eqnarray}}
\newcommand{\ea}{\end{eqnarray}}

\usepackage{color}
\usepackage[normalem]{ulem}

\usepackage{fancyhdr}
\begin{document}

\title{Exploring the propagation of relativistic quantum wavepackets 
in the trajectory-based formulation}

\author{Hung-Ming Tsai and Bill Poirier}

\address{Department of Chemistry and Biochemistry, and
         Department of Physics, \\
          Texas Tech University, Box 41061,
         Lubbock, Texas 79409-1061, USA}

\ead{Bill.Poirier@ttu.edu}

\begin{abstract}
In the context of nonrelativistic quantum mechanics, Gaussian wavepacket solutions
of the time-dependent Schr\"odinger equation provide useful physical insight.  
This is not the case for \emph{relativistic} quantum mechanics,
however, for which both the Klein-Gordon and Dirac wave equations result in strange
and counterintuitive wavepacket behaviors, even for free-particle Gaussians.
These behaviors include zitterbewegung and other interference effects.  
As a potential remedy, this paper explores 
a new trajectory-based formulation of quantum mechanics, in which the
wavefunction plays no role [\emph{Phys. Rev. X}, {\bf 4}, 040002 (2014)].  Quantum 
states are represented as ensembles of trajectories, whose mutual interaction is the source
of all quantum effects observed in nature---suggesting a ``many interacting worlds'' interpretation. 
It is shown that the relativistic generalization of the trajectory-based formulation
results in well-behaved free-particle Gaussian wavepacket solutions. In particular, probability 
density is positive and well-localized everywhere, and its spatial integral is conserved 
over time---in any inertial frame.  Finally, the ensemble-averaged wavepacket motion is along
a straight line path through spacetime.  In this manner, the pathologies of the wave-based 
relativistic quantum theory, as applied to wavepacket propagation, are avoided. 
\end{abstract}

\section{Introduction  \label{sec:intro}}

Quantum mechanics is generally regarded to be
a theory based on wavefunctions \cite{vonneumann,cohen-tannoudji,bohm}.  The quantum wavefunction, 
$\Psi$, is thought to be the fundamental mathematical representation of the state of a quantum system. 
As such, $\Psi$ has  always enjoyed a hallowed status, despite much historical 
and ongoing disagreement about its precise interpretation or physical significance
\cite{styer02,bohm52a,holland:1993bk,einstein35,ballentine70,home92,everett:1957pr,wheeler}.
Even ``alternative'' interpretations of quantum mechanics such as  Bohmian mechanics \cite{bohm52a,holland:1993bk}, 
which attribute physical reality to a ``quantum trajectory,'' nevertheless still adopt  a hybrid ontology 
in practice, wherein it is a combination of both the wavefunction {\em and} the quantum trajectory, \emph{together}, that  
specify the quantum state.

This article explores a fundamentally different kind of quantum theory, that makes no direct or indirect recourse 
to wavefunctions. It resembles Bohmian mechanics in that quantum trajectories are indeed employed.
However, unlike Bohmian mechanics, \emph{only} trajectories are used---the wave being replaced with a 
trajectory \emph{ensemble}, which thereby represents the quantum state
\cite{bouda:2003ijmpa,holland:2005ap,Poirier:2010zza,holland10,poirier11nowaveCCP6,Schiff:2012jcp,poirier12ODE,wiseman14prx,poirier14prx}.
The trajectory ensemble is continuous, with each individual member trajectory labeled by the parameter $C$.  The time evolution
of the trajectory ensemble, $x(t,C)$, is governed by some partial differential equation (PDE) in $(t,C)$ that replaces
the usual time-independent Schr\"odinger equation governing the $\Psi(t,x)$ evolution. All quantum effects
manifest as a dynamical interaction between neighboring trajectories in the ensemble---i.e., as partial derivatives in $C$. 
A new interpretation of quantum mechanics is thus also suggested by the new mathematics---what has come to be known as 
``many interacting worlds'' \cite{Poirier:2010zza,wiseman14prx,poirier14prx}. 

Interpretations notwithstanding, in the nonrelativistic context described above, the wave-based and (continuous) 
trajectory-based mathematics are equivalent;  thus, no new experimental predictions are proffered by the latter. 
On the other hand, a recent extension of the trajectory-based formulation to the \emph{relativistic} quantum 
regime \cite{Poirier:2012gz}---for single, massive, spin-zero, free particles, propagating on a flat Minkowski 
spacetime---does indeed provide new experimental predictions. However there are also other motivations for 
considering the relativistic extension of the trajectory theory.
 For example, the analogous wave-based relativistic quantum PDE for 
massive spin-zero particles---i.e. the Klein-Gordon equation---is riddled with foundational difficulties, when
interpreted as a single-particle theory. Most problematically, the temporal component of the four-current is 
not positive over all spacetime, and therefore cannot be interpreted as a probability density
 \cite{Ryder:1996bk,Wachter:2011bk}.

This  ``negative probability'' issue is mostly resolved when one moves to spin-1/2 particles such as electrons, for which 
the Dirac equation is used to describe the relativistic quantum dynamics. On the other hand, both the Klein-Gordon
and Dirac equations must contend with another difficulty, i.e. negative-energy solutions. Conceptually, one can address 
the negative-energy quantum states  somewhat satisfactorily---in familiar terms involving positrons, the ``Dirac sea,'' etc. 
In practice, however, negative-energy states lead to highly undesirable outcomes, that manifest even in that
simplest and most pedagogically useful of examples, the free-particle Gaussian wavepacket \cite{thaller,park12}. 
By far the most famous of these undesired outcomes is zitterbewegung---rapid oscillations of the position expectation value over time,
about a straight-line path through spacetime.  However, this is but one of several ``strange'' behaviors observed
in relativistic free-particle wavepacket dynamics, associated with interference between positive- and negative-energy
components.   These can be so severe that even 85+ years after they were first theoretically predicted, 
the physical origin and interpretation of these effects is still controversial, leaving even the 
standard probabilistic interpretation of the Dirac theory in doubt \cite{thaller,park12}.  A trajectory-based description
may help substantially, in at least two different ways.  First, it may admit an \emph{analytical} wavepacket solution; 
according to one author, much of the remaining controversy ``is all because there is no known analytical expression 
for a Dirac packet'' \cite{park12}.  Second, the relativistic quantum trajectory-based PDE \emph{has} no negative-energy
solutions, and is thereby possibly able to side-step the above difficulties, at least in principle. In any event, it is perhaps
telling that true zitterbewegung has yet to be observed experimentally \cite{park12}. 

In this paper, we analyze the trajectory-based dynamical PDEs for spin-zero relativistic quantum particles, as derived in 
Ref. \cite{Poirier:2012gz}. We also propagate those PDEs numerically, for various (initially) Gaussian wavepacket examples.
We begin in Sec.~\ref{sec:DE-3d}, by introducing the dynamical PDEs for a relativistic quantum particle in 
$(3+1)d$ spacetime. Then, in Sec.~\ref{sec:GWP-1d}, we reduce the spacetime dimension to $(1+1)d$ and derive
 dynamical equations specific to Gaussian wavepackets. Numerical solutions are also presented.
The symmetry properties of the dynamical PDEs, together with their numerical solutions, are then investigated in 
Sec.~\ref{sec:SI}, with respect to a set of scale transformations that preserves the equations of motion. The predicted scale
invariance is verified both analytically and numerically.  In this section, also, we examine the spacetime dependence
of the spatial metric, which provides important dynamical information---e.g.,  as pertains  to the relativistic restriction on
wavepacket broadening.   

Whereas an analytical solution may someday be tenable for the relativistic free-particle Gaussian wavepacket, 
at present we must rely on numerical solutions. Sec.~\ref{sec:NUG} therefore addresses numerical issues, 
especially the issue of how to systematically characterize and improve the convergence accuracy of the numerical
trajectory ensemble solutions. This is more difficult than for many standard numerical analyses, owing to the fact that
the true boundary conditions are unknown. As a consequence, numerical instabilities are sometimes encountered
under certain conditions---although we provide one simple strategy for ameliorating this difficulty, involving nonuniform
grids, which seems to work rather well.

The analysis up to this point in the paper pertains only to ``stationary'' Gaussian wavepackets, which  remain centered
at $x=0$ throughout time. In Sec.~\ref{sec:LBF}, we generalize everything for arbitrary moving free-particle Gaussian 
wavepackets, by simply applying Lorentz boosts to the stationary wavepacket solutions.  In doing so, we confirm the 
Lorentz invariance of the dynamical PDEs and their numerical solutions, as well as the associated flux four-vectors. 
We also examine the probability density in both inertial frames,
confirming that this quantity---the temporal component of the flux four-vector---is indeed positive throughout spacetime, and in all 
inertial frames.  Finally, we confirm probability conservation by explicitly integrating the probability density 
over all space, and showing that the result is conserved over time---in any inertial frame.  Thus, the trajectory-based
treatment of relativistic free-particle Gaussian wavepackets does indeed appear to avoid the undesirable features
of the corresponding wave-based approaches. 


\section{Dynamical equations in $(3+1)d$ spacetime \label{sec:DE-3d}}

The following is a very brief overview of the derivation provided in Ref.~\cite{Poirier:2012gz}, which should be consulted for further details. 
In the trajectory-based formulation for a single spin-free relativistic quantum particle in flat $(3+1)d$ spacetime, a quantum state is represented as an ensemble of quantum trajectories, $x^\alpha (X^\mu)$. 
The \emph{dependent} field variables are the inertial or extrinsic coordinates $x^\alpha=(ct, \bm{x}),$ whereas the independent variables are the natural or intrinsic coordinates  $X^\mu = (c\lambda, \bm{C} )$. The quantity $\lambda$ is a global  time-like parameter called the ``ensemble time,'' whereas the three space-like $\bm{C}$ coordinates label the individual quantum trajectories in the ensemble. Contours 
of the former, expressed in inertial coordinates, represent ``simultaneity submanifolds''---i.e., global space-like-separated sets of events that occur simultaneously for the quantum particle.  Contours of the latter---or rather, their mutual intersections---define the quantum trajectories themselves.
These foliate spacetime, and may be interpreted as a continuous spectrum of ``worlds,'' in each of which resides a single ``copy'' of the particle.  
Both $\lambda$ and $\bm{C}$ may be separately reparametrized, in arbitrary fashion.  

We denote the metric tensor of the natural coordinate system as $g^{\mu \nu}$, and that of the inertial coordinates as $\eta^{\alpha \beta}$. These are related via the transformation,
\ba
{g_{\mu \nu }} = {\eta _{\alpha \beta }}\frac{{\partial {x^\alpha }}}{{\partial {X^\mu }}}\frac{{\partial {x^\beta }}}{{\partial {X^\nu }}}.  \label{eq:g_metric}
\ea
In matrix form, the inertial metric tensor is the familiar $\tilde \eta = \mathrm{diag}(-1,1,1,1)$. For the natural coordinates, we impose
local orthogonality between infinitesimal displacements in $\lambda$ and $\bm{C}$. This results in a block-diagonal metric tensor,  i.e.
\ba
\tilde g = \left( {\begin{array}{cc}
   g_{00} & 0 \\
   0 & \tilde \gamma  \\
  \end{array} } \right),  \label{eq:g_matrix_form}
\ea
where the ``spatial metric'' $\tilde \gamma$ denotes the $3 \times 3$ spatial block of $\tilde g$.  We further define
\ba
g= \det \tilde g \qquad ;  \qquad \gamma=\det\tilde{\gamma}.   \label{eq:det_g_det_gamma}
\ea
Finally, we note that $g_{00} = -\left( d \tau/d \lambda  \right)^2$, where $\tau$ is the usual 
proper time, defined  via
\ba
d \tau^2 = -\frac{1}{c^2} \eta_{\alpha \beta} dx^\alpha dx^\beta.  \label{eq:dtau_definition}
\ea

Probability conservation along individual quantum trajectories is presumed \cite{Poirier:2012gz}, as a result of which
the spatial scalar probability density on $\bm{C}$-space, $f(\bm{C})$, must be independent of  $\lambda$.  In
addition to this scalar quantity, there is also the {\em flux four-vector}, 
\ba
        j^\mu = f({\bf C})\, {dX^\mu\over d\lambda} =  (c f({\bf C}),0,0,0),
        \label{genflux}
\ea
which transforms as a contravariant vector density of weight  $W\!=\!-1$.  Due to a fortuitous cancellation of 
Christoffel symbols, $j^\mu$  satisfies the following covariant continuity equation, 
\ba
        \partial_\mu\, j^\mu = 0,
        \label{continuity}
\ea
in \emph{all} coordinate frames. In an inertial frame, in particular, this leads to the familiar
integrated form of probability conservation,  
\ba
        \int j^0(x^\alpha)\, d^3 x  = {\rm const}.
        \label{Pcons}
\ea
It is noteworthy that  $j^0$ can be interpreted as a true \emph{probability} density 
(apart from a scaling factor), as in nonrelativistic quantum mechanics. 
The analogous Klein-Gordon $j^0$ quantity, for example, can change 
sign, and must therefore be interpreted as a \emph{charge} density \cite{Ryder:1996bk, Wachter:2011bk}.

In addition to its usual probabilistic role,  $f({\bf C})$ also plays an indirect \emph{dynamical} role in 
the quantum trajectory theory.  The quantum potential $Q$ and the quantum force $f^\alpha$ are given by
\ba
Q &=&  - \frac{{{\hbar ^2}}}{{2m}}{\gamma ^{ - 1/4}f^{-1/2}}\frac{\partial }{{\partial {C^i}}}\left[ {{\gamma ^{1/2}}{\gamma ^{ij}}\frac{\partial }{{\partial {C^j}}}{ \left( f^{1/2} \gamma   ^{ - 1/4} \right)   }} \right],  \label{QPOT}\\
{f^\alpha } &=& -\eta^{\alpha \beta} \frac{\pa C^i}{ \pa x^\beta} \frac{\pa Q}{\pa C^i} = - \frac{{\partial {x^\alpha }}}{{\partial {C^i}}}{\gamma ^{ij}}\frac{{\partial Q}}{{\partial {C^j}}}, \label{QF}
\ea
where $\gamma^{ij}$ is the matrix form of $\tilde{\gamma}^{-1}$.  The above equations are valid for any 
parametrization of $\lambda$ and ${\bf C}$.  On the other hand, we can make a special choice for $\lambda$,
known as the ``ensemble proper time,'' $\mathcal{T}$, which reduces to $\tau$ when $Q=0$. More generally~\cite{Poirier:2012gz}, the
relation is 
\ba
\frac{d \tau}{d \mathcal{T}} = \exp \left( -\frac{Q}{mc^2}  \right).
\ea 
This leads, finally, to the dynamical equations of motion for a free particle:
\ba \frac{\pa^2 x^\alpha }{\pa
\mathcal{T}^2} = \exp\left( \frac{-2Q}{mc^2} \right) \frac{f^\alpha}{m} -\frac{1}{mc^2} \frac{\pa Q}{\pa \mathcal{T}} \frac{\pa x^\alpha}{\pa \mathcal{T}}.  \label{eq:rel_PDE-3d}
\ea
Solving  Eq.~(\ref{eq:rel_PDE-3d}) requires specification of the initial conditions,
$x^\alpha_0 = x^\alpha(0,{\bf C})$ and $(\partial x^\alpha /  \mathcal{T}) (0,{\bf C})$. 

Note that Eq.~(\ref{eq:rel_PDE-3d}) is second order in $\mathcal{T}$ and fourth order in $\bf C$---in exact analogy with the time and space orders of the Schr\"odinger PDE (the order doubling is due to the fact that the field quantity is in this case real- rather
than complex-valued).  Thus, \emph{natural} time and \emph{natural} space are not treated on an equal footing in this relativistic 
picture---but then again, neither should they be. In contrast, the \emph{inertial} coordinates, $x^\alpha$ \emph{are} treated equally, in the
sense that Eq.~(\ref{eq:rel_PDE-3d}) is Lorentz-invariant.  In this manner---i.e., by separating natural from inertial coordinates---we 
are able to avoid the relativistic quantum ``paradox'' that, for example, led Dirac to consider multiple wave components.  Moreover, the 
negative energy solutions that one finds with Klein-Gordon \cite{Ryder:1996bk, Schweber:1961bk, Gross:1999bk} or Dirac \cite{Ryder:1996bk, Schweber:1961bk, Gross:1999bk} also go away, because one finds that positive-negative
pairs of wavefunction solutions correspond to a single trajectory ensemble solution $x^\alpha (X^\mu)$  \cite{Poirier:2012gz}.


\section{Gaussian wavepacket solutions in $(1+1)d$ spacetime \label{sec:GWP-1d}}

Some analytical and numerical solutions of Eq.~(\ref{eq:rel_PDE-3d}) are provided in Ref.~\cite{Poirier:2012gz}, for 
 $(1+1)d$ spacetime. In this reduced-dimensional context, there is but a single inertial (natural) spatial coordinate, $x=x^1$ ($C=C^1$), 
 and the dynamical equations simplify as a consequence of
\ba
\frac{1}{\gamma^{ij}} = \gamma_{ij} = \gamma  = 
\left(  \frac{\pa x}{\pa C} \right)^2- c^2 \left( \frac{\pa t}{\pa C}  \right)^2.   \label{eq:gamma-1d}
\ea

For the  remainder of this article, we will focus on  $(1+1)d$ spacetime, and specifically on the free particle
propagation of the relativistic Gaussian wavepacket.  In nonrelativistic quantum mechanics, the Gaussian
wavepacket is well known to disperse while moving in a straight line---yet retains a perfect Gaussian shape 
at all times \cite{cohen-tannoudji}. The wavepacket  spreading is therefore linear. This behavior cannot also 
characterize the relativistic regime, for it would imply the existence of faster-than-light quantum trajectories 
in the Gaussian tails, which are precluded by the $(1+1)d$ Eq.~(\ref{eq:rel_PDE-3d}).  At best, the relativistic
wavepacket can be truly Gaussian at a single time only.  Moreover, one has to resolve which 
``time'' is being referred to, i.e. $\mathcal{T}$ or $t$. 

Both issues are resolved by considering coherent-state or minimum-uncertainty Gaussians, which exhibit 
uniform flux/velocity throughout $x$. All nonrelativistic free-particle Gaussian wavepackets evolve into
a coherent state at exactly one point in time during their lifecycles.  By setting this time equal to zero, and 
transforming to a frame in which the uniform velocity is zero, all quantum trajectories in the ensemble 
become instantaneously stationary. The relativistic and nonrelativistic descriptions must therefore agree 
in that initial instant. 
In particular, the $t=0$ and $\mathcal{T}=0$ initial conditions become identical.  Our ansatz for a 
free-particle relativistic Gaussian wavepacket therefore becomes a solution of $(1+1)d$ Eq.~(\ref{eq:rel_PDE-3d}) 
whose initial conditions satisfy:
\ba
t_0 (C) = 0 \quad ; \quad  x_0(C) = C  \quad ;  \quad  \frac{{\partial x}}{{\partial \mathcal{T}}}\left( {0,C} \right) = 0  \label{eq:IC}, 
\qquad \mbox{with  } f(C)= \exp\left( -a C^2 \right). 
\ea
Note that Eq.~(\ref{eq:IC}) implies a specific parametrization for $C$---i.e., for any given trajectory, $C$ is the 
initial $x$ value, $x_0$  \cite{Poirier:2010zza,Poirier:2012gz}.  Likewise, our earlier specification of the natural time coordinate, 
$\lambda = \mathcal{T}$, implies the following as the one remaining initial condition:
\ba
\frac{{\partial t}}{{\partial \mathcal{T}}}\left( {0,C} \right) =
 \exp \left[ { - \frac{{{Q_0}\left( C \right)}}{{m{c^2}}}} \right] = \exp \left[ {\frac{1}{2}{{\left( {\frac{\hbar }{{mc}}} \right)}^2}\left( {{a^2}{C^2} - a} \right)} \right].
 \label{eq:IC-2} 
\ea

\begin{figure}
\includegraphics[width=0.45\textwidth]{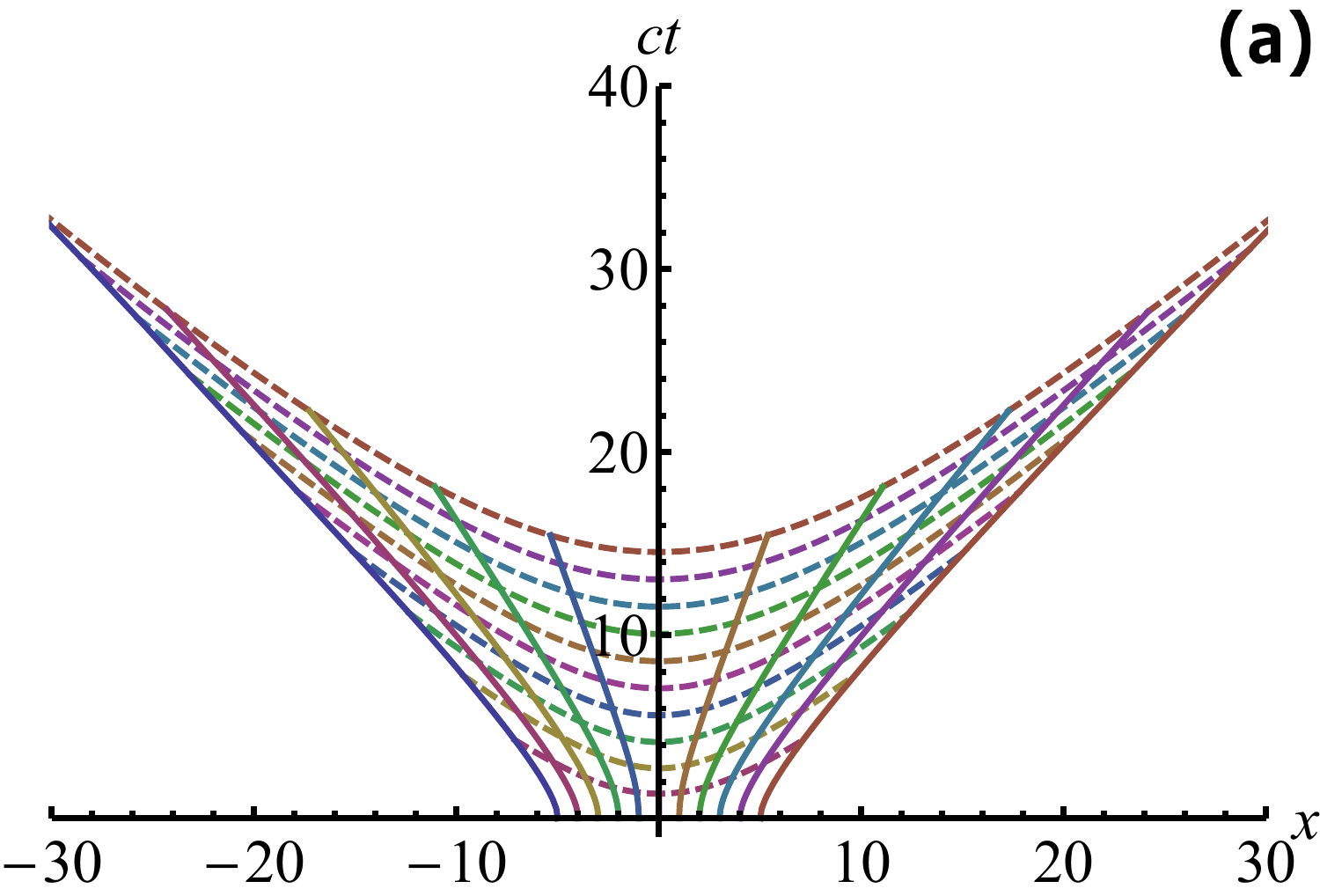}
\includegraphics[width=0.45\textwidth]{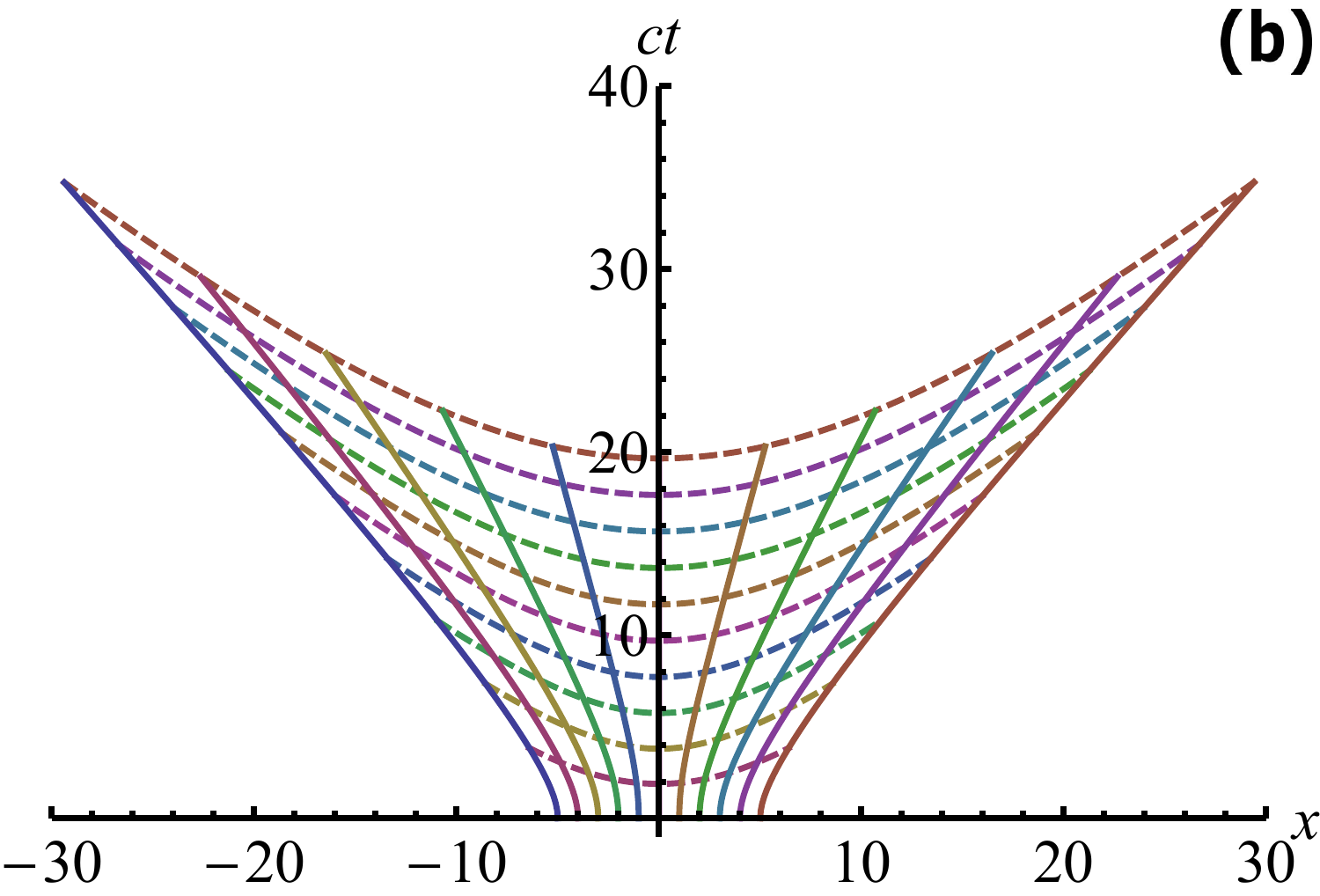}
\includegraphics[width=0.45\textwidth]{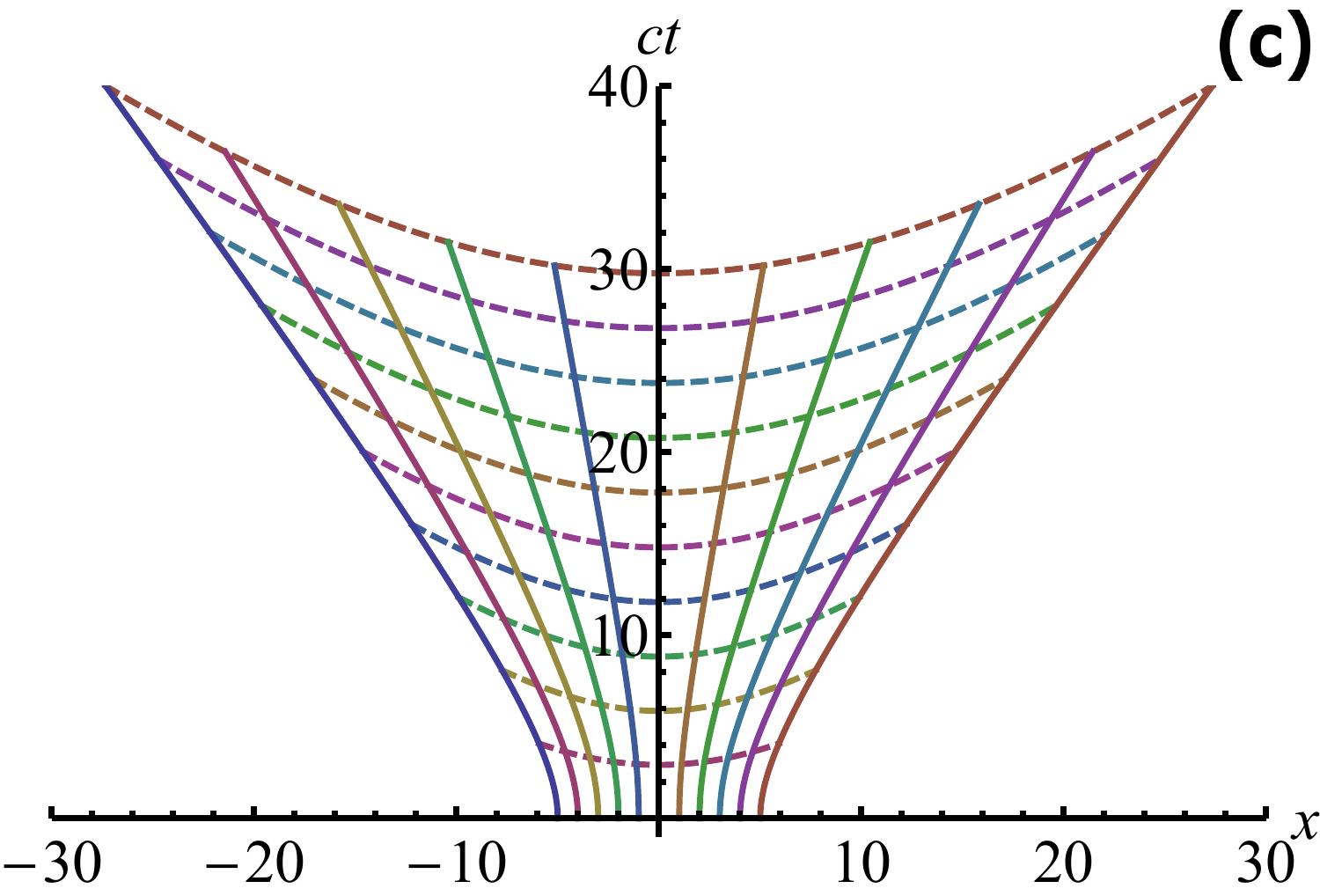}\hfill
\includegraphics[width=0.45\textwidth]{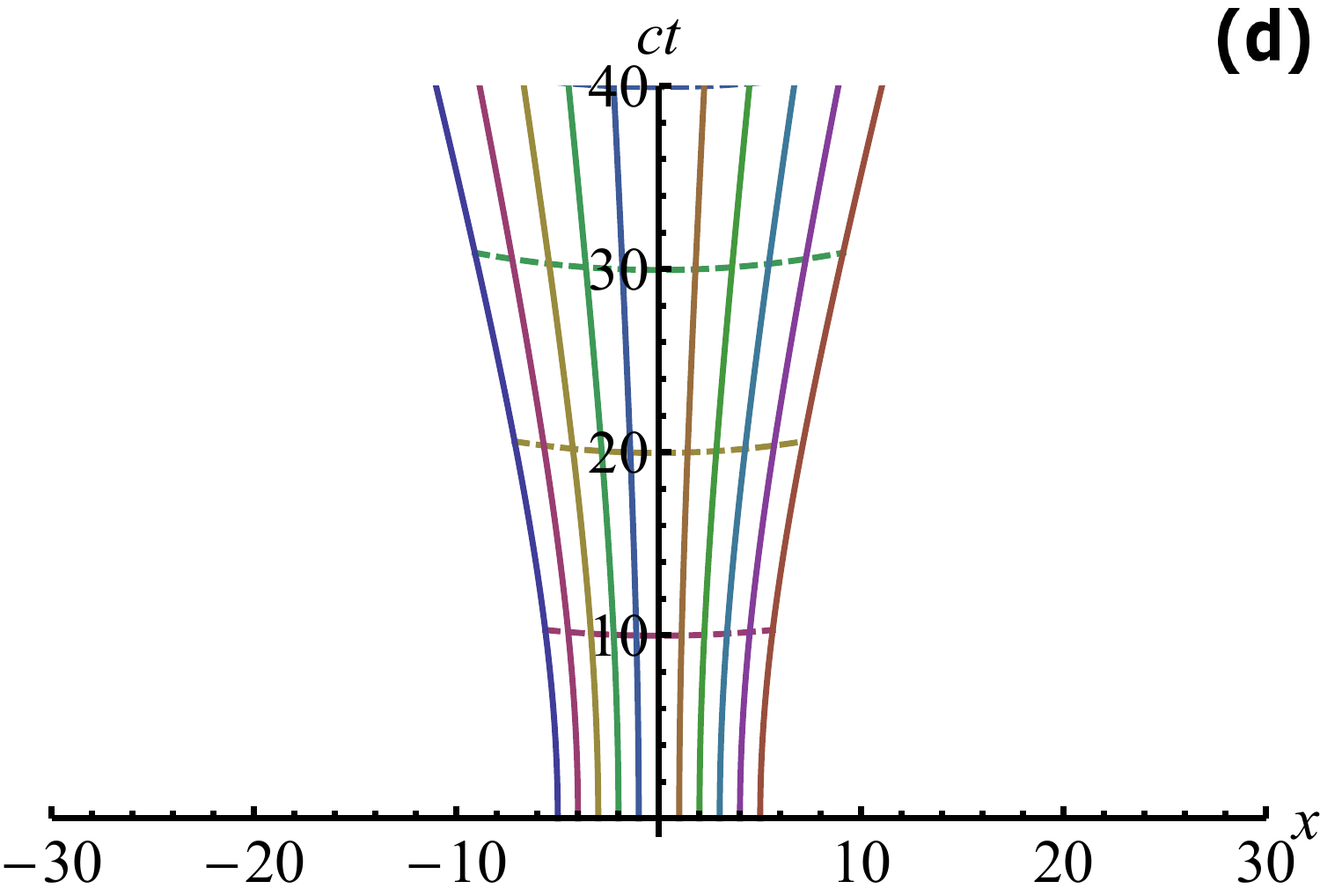}
\caption{Quantum trajectories (solid curves) and simultaneity submanifolds (dashed curves) 
for the relativistic free-particle Gaussian wavepacket of Eq.~(\ref{eq:IC}), 
for $a=1/2$, $\hbar = m=1$, and several choices of the speed of light, $c$: 
(a) $c=1.5$; (b) $c=2.0$;  (c) $c=3.0$;  (d) $c=10$.}
\label{fig:GWP-1}
\end{figure}

In Eq.~(\ref{eq:IC-2}) above, $Q_0(C) = Q(0, C)$ is the initial quantum potential. More generally, 
Eqs.~(\ref{QPOT}) and (\ref{QF}) reduce to the following, for $(1+1)d$ spacetime with 
the $f(\bf{C})$ choice of Eq.~(\ref{eq:IC}):
\ba
Q &=& - \frac{\hbar^2}{2m} \frac{e^{\frac{1}{2} a C^2}}{\gamma^{1/4}}  \frac{\pa}{\pa C} \left[  \frac{1}{\gamma^{1/2}} \frac{\pa}{\pa C}  \left(e^{-\frac{1}{2} a C^2}   \gamma^{-1/4} \right)\right];  \label{eq:Q-2} \\
f^0 &=& -  \frac{c}{\gamma }\frac{\pa t }{\pa C} \frac{\pa Q }{\pa C} \quad ; \quad f^1 =  -  \frac{1}{\gamma }\frac{\pa x }{\pa C} \frac{\pa Q }{\pa C}.   \label{eq:force-2}
\ea
The resultant dynamical equations become
\ba
-\frac{1}{m} \frac{1}{\gamma} \frac{\pa t}{\pa C  }  \frac{\pa Q}{\pa C} &=& \exp \left( \frac{Q}{mc^2} \right) \frac{\pa}{\pa \mathcal{T}} \left[ \exp \left( \frac{Q}{mc^2}   \right)  \frac{\pa t}{\pa \mathcal{T}  }  \right], \nonumber \\
-\frac{1}{m} \frac{1}{\gamma} \frac{\pa x}{\pa C  }  \frac{\pa Q}{\pa C} &=& \exp \left( \frac{Q}{mc^2} \right) \frac{\pa}{\pa \mathcal{T}} \left[ \exp \left( \frac{Q}{mc^2}   \right)  \frac{\pa x}{\pa \mathcal{T}  }  \right].  \label{eq:PDE_two}
\ea 

To date, analytical solutions of Eq.~(\ref{eq:PDE_two}) for $t(\mathcal{T},C)$ and $x(\mathcal{T},C)$ have 
eluded us.  However, we have obtained numerical solutions, for  $a=1/2$, $\hbar = m=1$, and various values 
of the speed of light $c$, as presented in Fig.~\ref{fig:GWP-1}.  Wavepacket spreading is evident,
as in the nonrelativistic case. The quantum trajectories approach the light cone faster in the ultrarelativistic 
limit of small $c$, as expected.  The simultaneity submanifolds also become increasingly distorted in this limit,  
whereas in the nonrelativistic (large $c$) limit,  they approach horizontal lines (i.e., contours of $t$).


\section{Scale invariance of the dynamical solutions, and the spatial metric \label{sec:SI}}

The PDE of Eq.~(\ref{eq:PDE_two}) governs the time evolution of the quantum trajectory ensemble for the 
relativistic free-particle Gaussian wavepacket. Rescaling the system parameters of 
this PDE alters its solutions. However, it is possible to rescale the parameters and dynamical variables together, in such a 
way that the solutions remain invariant.
Consider the double scale transformation defined in terms of the two parameters, $\zeta$ and $\eta$, with $\zeta$ associated
with $c$, and $\eta$ with the independent and dependent variables, as follows:
\ba
 c \to \zeta c \quad ; \quad  X^\mu \to \eta X^\mu \quad ; \quad  x^\alpha \to \eta x^\alpha.
\ea
\vskip-0.7cm
\ba
\mathcal{T} \to \frac{\eta}{\zeta} \mathcal{T} \quad ;  \quad C \to \eta C \quad ; \quad  t \to \frac{\eta}{\zeta} t \quad ;  \quad x \to \eta x.  
\label{eq:coordinates}
\ea
For the quantum trajectory ensemble solution to remain invariant, Eq.~(\ref{eq:IC}) implies that $f(C)$ must  
also remain invariant. 
Under the rescaling $C \to \eta C$, this requires $a \to (1 / \eta^2)a$. 
More generally, the  invariance of the solution under $C \to \eta C$ rescaling implies:
\ba
\gamma \to \gamma \,\,\,\mbox{from Eqs.~(\ref{eq:gamma-1d}) and (\ref{eq:coordinates})} \qquad; 
\qquad  \frac{Q}{mc^2} \to \frac{Q}{mc^2}\,\,\,\mbox{from Eq.~(\ref{eq:PDE_two})} \label{eq:exp_Q_mc2}
\ea
\vskip-0.7cm
\ba
\left(\frac{\hbar}{mc}\right)^2 \to \eta^2 \left(\frac{\hbar}{mc}\right)^2  \,\,\,\mbox{from Eqs.~(\ref{QPOT}) and (\ref{eq:exp_Q_mc2}).} 
\label{eq:hbar_mc2}
\ea
Now we consider the invariance of the solution under  $c \to \zeta c$, which implies:
\ba
\frac{\hbar}{m} \to \eta \zeta \left(\frac{\hbar}{m}\right)\,\,\,\mbox{from Eq.~(\ref{eq:hbar_mc2})} \qquad; 
\qquad  \frac{Q}{m} \to \zeta^2 \frac{Q}{m}\,\,\,\mbox{from Eq.~(\ref{eq:exp_Q_mc2})}  \label{eq:Q_m}
\ea
\vskip-0.7cm
\ba
\frac{f^\alpha}{m} \to \frac{\zeta^2}{\eta} \left(\frac{f^\alpha}{m}\right) \,\,\,\mbox{from Eqs.~(\ref{eq:force-2}), (\ref{eq:coordinates}) and (\ref{eq:Q_m})}. 
\ea

\begin{figure}
\includegraphics[width=0.45\textwidth]{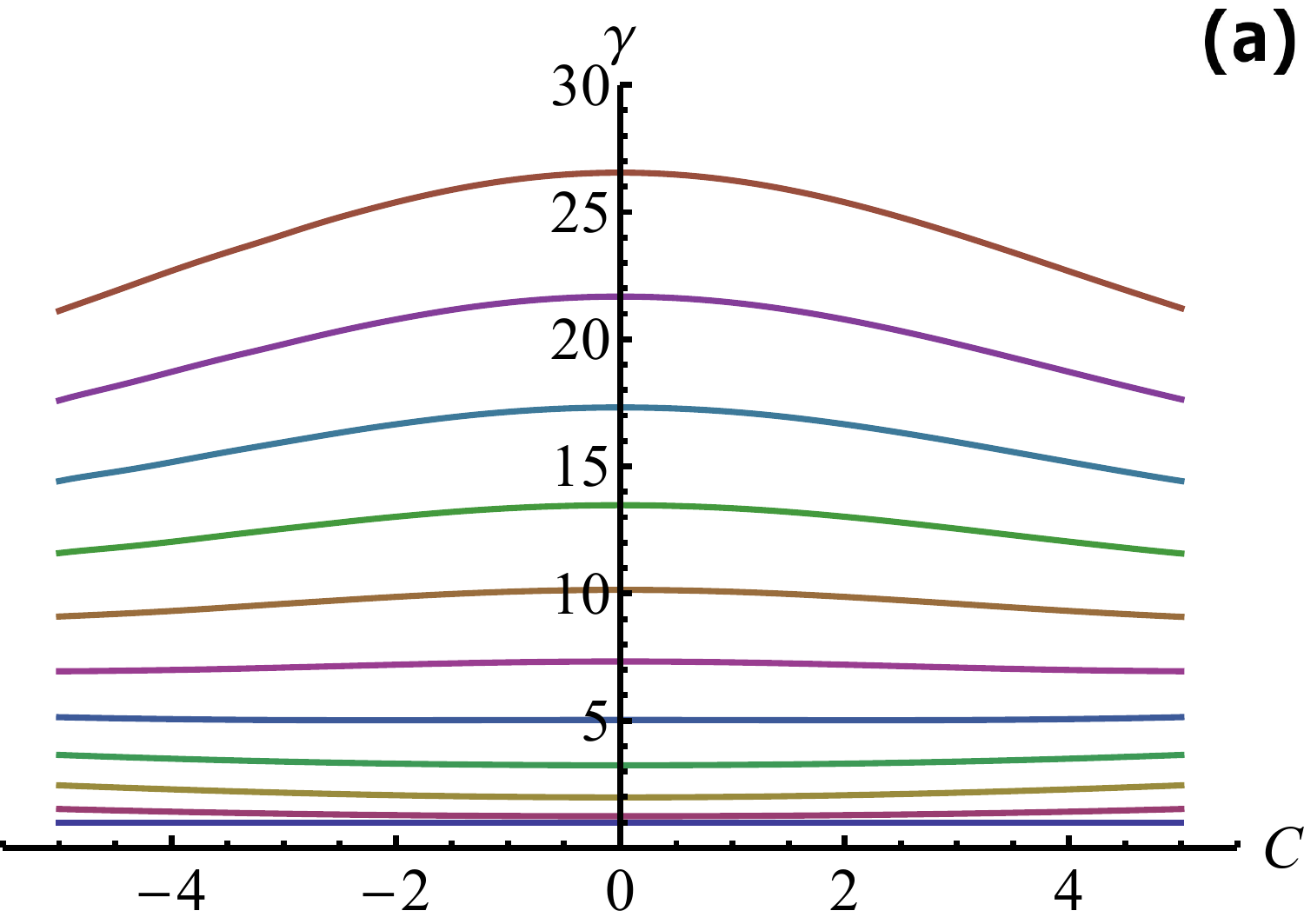} \hfill
\includegraphics[width=0.45\textwidth]{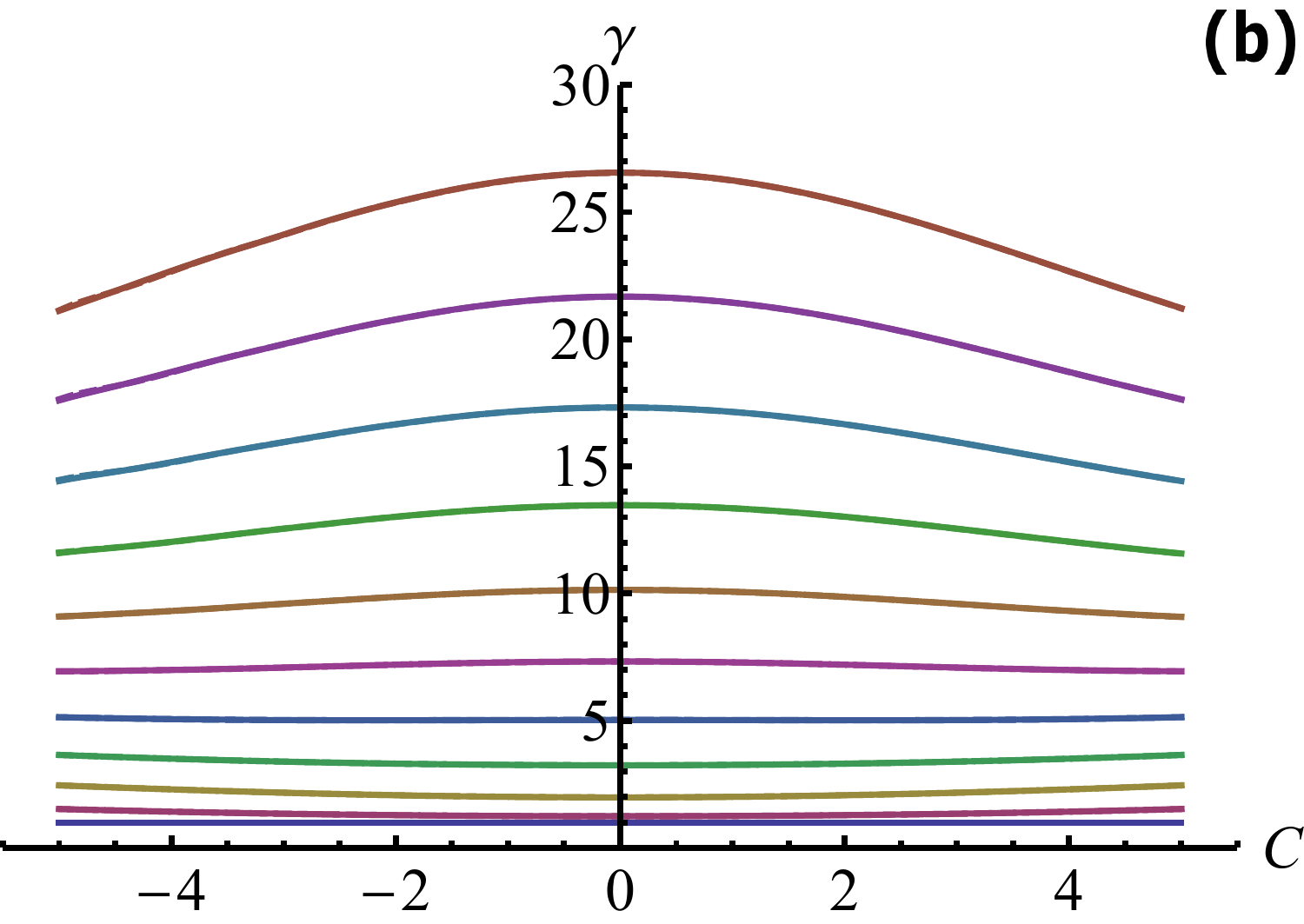}
\caption{Spatial metric $\gamma(\mathcal{T},C)$ for $(1+1)d$ relativistic free-particle 
Gaussian wavepackets. Each curve represents $\gamma(C)$ for fixed $\mathcal{T}=\{0,1,2 \ldots,10\}$, 
with increasing $\mathcal{T}$ values corresponding to increasing $\gamma$:
(a) $(a=1/2, \hbar = 1, c=3)$ solution; (b) comparison between $(a=1/2, \hbar = 1, c=3)$ and
$(a=1/5, \hbar = 5 \sqrt{10}/3, c=10)$ solutions, which are numerically nearly indistinguishable.}
\label{fig:SI-1}
\end{figure}

Finally, we examine the rescaling of Eq.~(\ref{eq:PDE_two}) directly.  Under the simultaneous
rescaling $C \to \eta C$ and  $c \to \zeta c$, the left hand side of Eq.~(\ref{eq:PDE_two}) is found to 
rescale as $\zeta^2/\eta$. The right hand side also rescales as  $\zeta^2/\eta$---thus implying
scale invariance for the free-particle Gaussian wavepacket evolution.  More generally, similar arguments
are also found to hold for Eq.~(\ref{eq:rel_PDE-3d}). The scale invariance has also been verified numerically. 
In Fig.~\ref{fig:SI-1}, we compare  $\gamma(\mathcal{T},C)$ for the $(a\! =\! 1/2, \hbar\! =\!  m\! =\!  1, c\! =\! 3)$ Gaussian 
wavepacket solution vs. that for $(a\! =\! 1/5, \hbar\!  =  5 \sqrt{10}/3, m\! =\! 1, c\! =\! 10)$---corresponding to 
the $(\zeta \! =\!  10/3, \eta\!  = \! \sqrt{5/2})$ rescaling.  We observe that all of the 
fixed-$\mathcal{T}$ $\gamma(C)$ curves are numerically nearly indistinguishable. 

Since $\gamma$ in fact represents the spatial metric, Fig.~\ref{fig:SI-1} provides useful insight
into the wavepacket dynamics. The primary feature evident in the plots  is that  
the $\gamma(C)$ curves increase in magnitude with increasing $\mathcal{T}$---an
indication of wavepacket broadening. Note, however, that this increase is not uniform across 
$C$---i.e.,  the curves are not horizontal lines.  If they were, this 
would imply preservation of the Gaussian form over time, which only occurs in the 
nonrelativistic limit. Conversely, the curvature or bowing of these curves---which for this example, 
is seen to be rather pronounced---is an indication of relativistic quantum 
dynamical effects.

Another interesting feature may also be observed in Fig.~\ref{fig:SI-1} , 
which is that the direction of the curvature changes over time.
The early curves bow upward, whereas at later times, the 
curves bow downward.  This can be explained as a competition
between the dynamical influences of  $Q$ and $f^\alpha$.  At early times,
all trajectories are ``moving'' in unison; their acceleration has not
yet had a chance to manifest as a fanning out of velocities.  However, even at
$\mathcal{T}\!\!=\!0$, the local proper time evolves at a higher rate towards the exterior fringes 
of the ensemble (i.e., towards larger $|C|$)---in accord with the relativistic quantum 
\emph{time compression} effect \cite{Poirier:2012gz}. The early simultaneity submanifolds thus curve
away from the $x$ axis, resulting in larger intervals between nearby trajectories that lie further 
from $C\!\!=\!0$---and thus, in upward-bowing $\gamma(C)$ curves.  Over time, acceleration 
gives rise to trajectory fanout, as discussed.  However, as the velocities of the exterior  
trajectories reach the order of the speed of light, further broadening of the wavepacket  is 
relativistically hindered, and so the resultant $\gamma(C)$ curves bow downward, rather 
than upward.   


\section{Numerical details \label{sec:NUG}}

From a numerical standpoint, the solution of Eq.~(\ref{eq:PDE_two}) offers important advantages over
both conventional, $x$-grid-based Crank-Nicholson propagation of $\Psi$, and
traditional quantum trajectory methods \cite{wyatt}.  Briefly, one has the
simultaneous advantages of \emph{both} a regular grid (in $C$) \emph{and} probability-conserving
trajectories (in $x$) \cite{Poirier:2010zza}.  There are nevertheless some nontrivial numerical issues,
stemming from the fact that the true boundary conditions are unknown---unlike for $\Psi$ propagation,
for which Dirichlet boundary conditions are in effect.  In practice, as a result of the small asymptotic
density $f(C)$, the less significant boundary regions seem to have very little effect on the far more important 
interior.  Consequently, one can choose any reasonable boundary condition---or none at all, as is
the case with most of our \emph{Mathematica} calculations (performed using {\bf NDSolve})---and still obtain
excellent results.  On the other hand, there are also situations when this boundary condition issue can cause
numerical instabilities. This is an ongoing area of investigation, though we discuss one simple 
example and remedy here.

\begin{figure}
\includegraphics[width=0.45\textwidth]{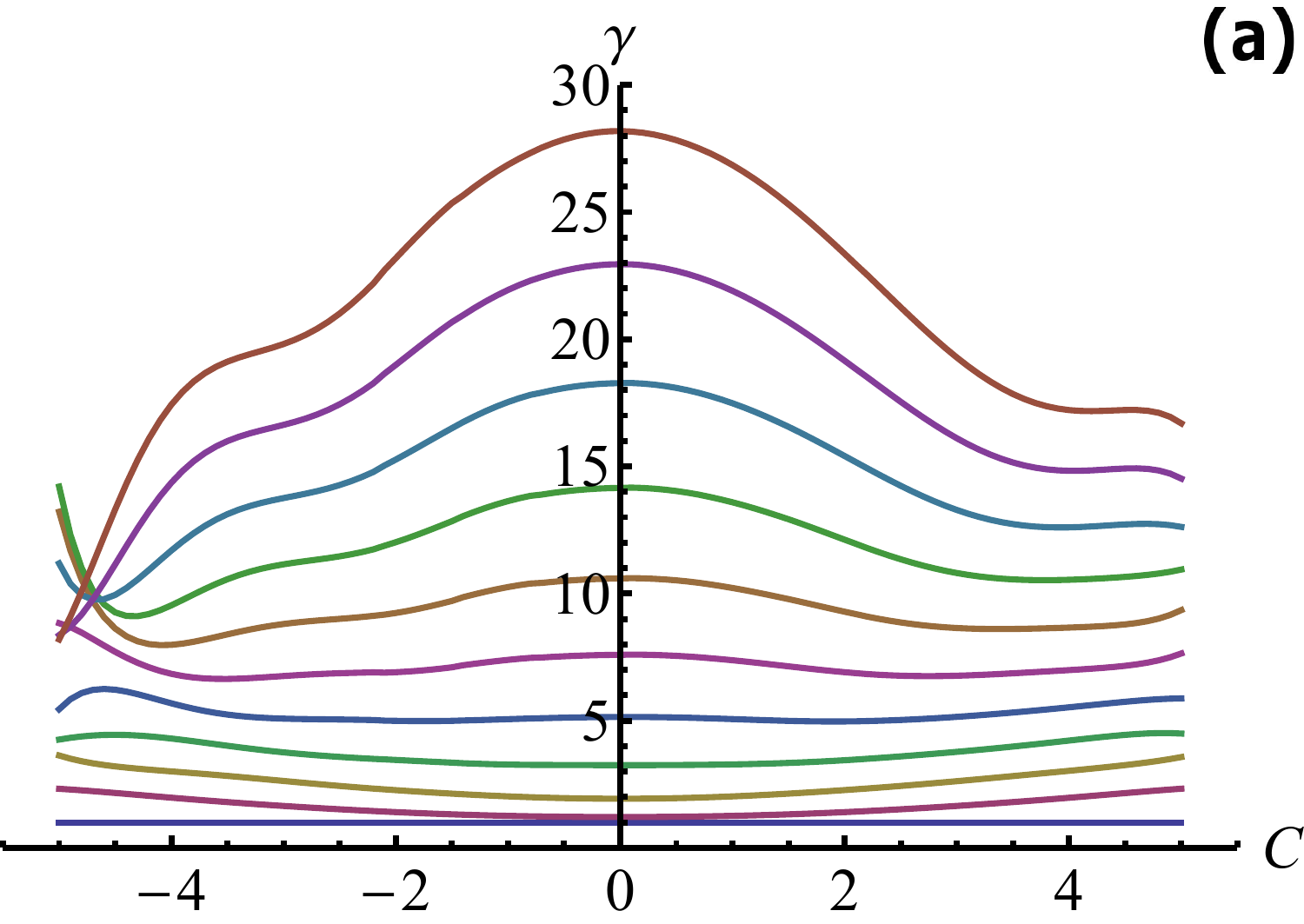}\hfill
\includegraphics[width=0.45\textwidth]{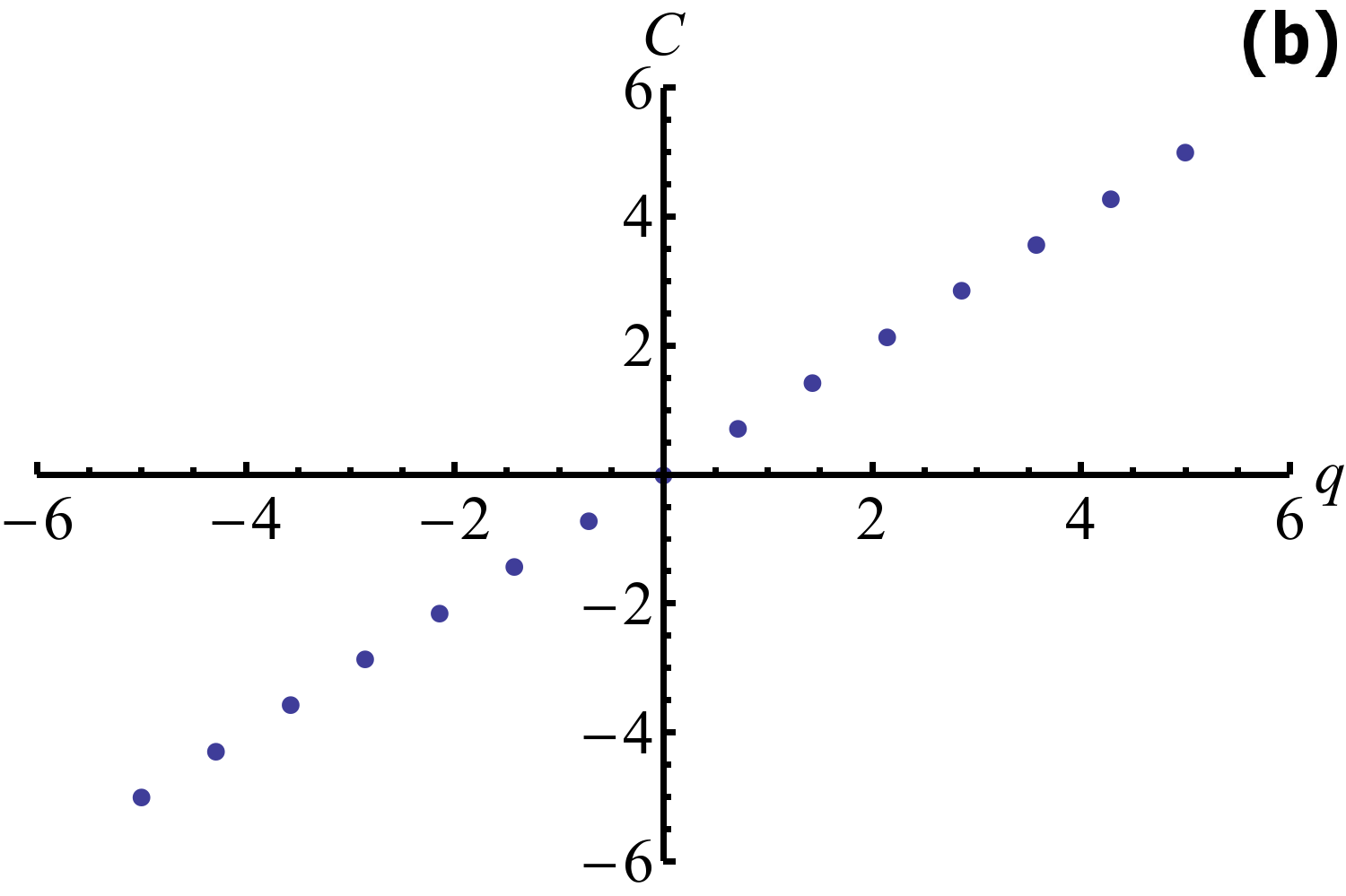}
\caption{Calculations with a uniformly-spaced $C$ grid: (a) numerically computed $\gamma(\mathcal{T},C)$,
as per Fig.~\ref{fig:SI-1}, but for $c=1.5$; (b) the $N_g=15$ grid points themselves, uniformly spaced over $-5 \le C\le 5$.}
\label{fig:UG-1}
\end{figure}

For the present Gaussian wavepacket calculations, errors and instabilities tend to increase 
when the numerical grid is extended over a larger $C$ domain, and also in the ultrarelativistic
limit.  Figure~\ref{fig:UG-1} presents $\gamma(C)$ curves for the $c=1.5$ example of Fig.~\ref{fig:GWP-1}(a),
as computed using a simple uniformly-spaced $C$ grid of $N_g=15$ points, over the domain 
$-C_{\rm{max}} \le C \le C_{\rm{max}}$, with  $C_{\rm{max}}=5$. 
The lack of symmetry, curve-crossing, and sharp changes are all evidence of numerical inaccuracy in the boundary
regions, although the interior trajectories where probability is significant are still fairly accurate.  Simply increasing
the number of grid points, however, without changing the $C$ domain, does not lead to improved accuracy,
and actually makes the boundary behavior worse.  The same is true if $C_{\rm{max}}$ is increased, using the same
grid density.

Based on these observations, we instead adopt a \emph{non}uniform
$C$ grid, wherein a higher density of grid points is used in the more important interior region. This is achieved
via the map,
\ba
C(q) = A \tanh^{-1} (\beta q),   \label{eq:C_q}
\ea
where grid points are now uniformly distributed over $-q_{\max} \le q \le q_{\max}$. The quantity
$A = C_{\max}/ \tanh^{-1}(\beta q_{\max })$, and $\beta$ is the nonuniformity parameter.  The nonuniform
grid greatly improves asymptotic behavior, thereby allowing for higher grid densities and more accurate 
calculations in the interior.  

\begin{figure}
\includegraphics[width=0.45\textwidth]{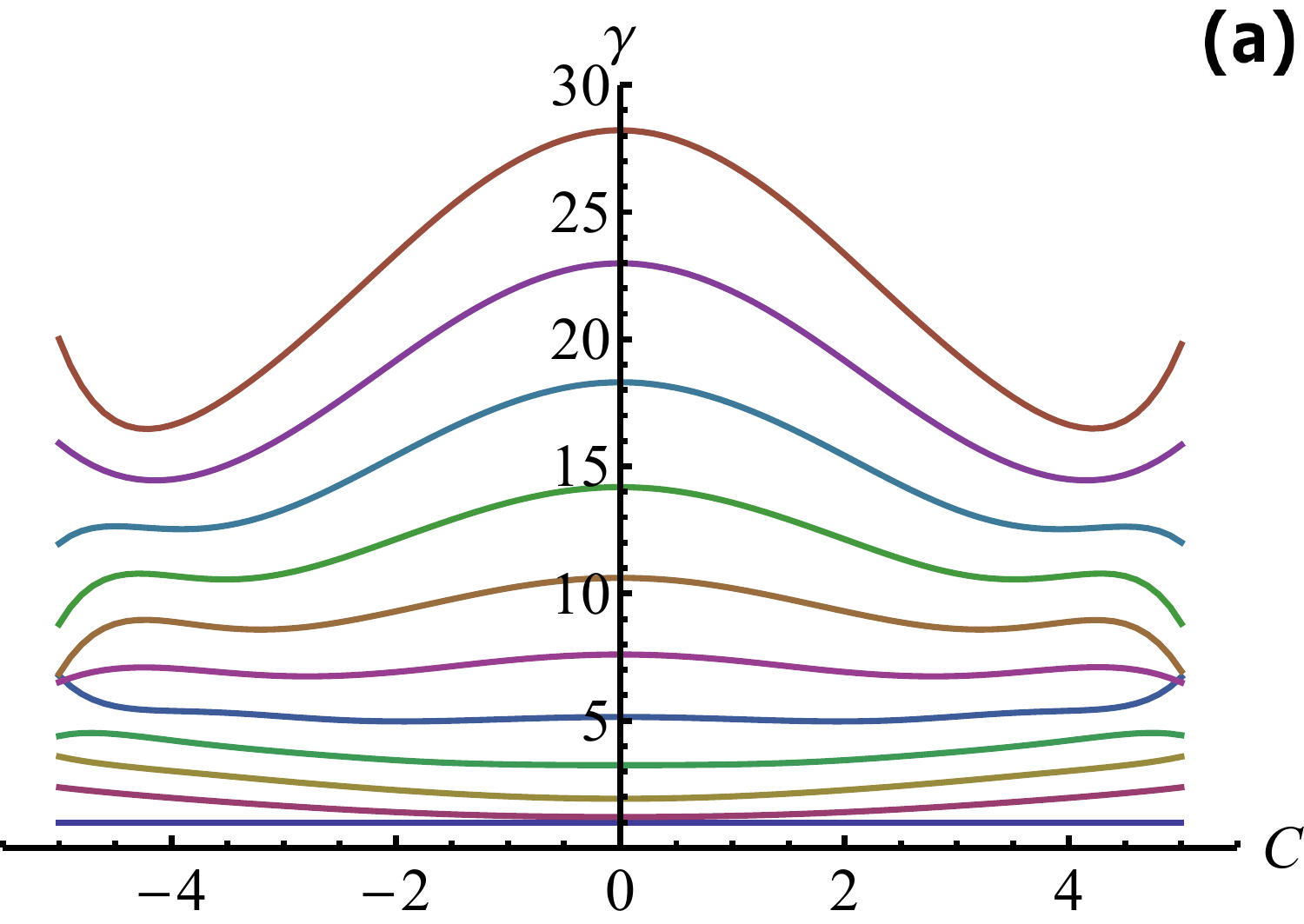}
\includegraphics[width=0.45\textwidth]{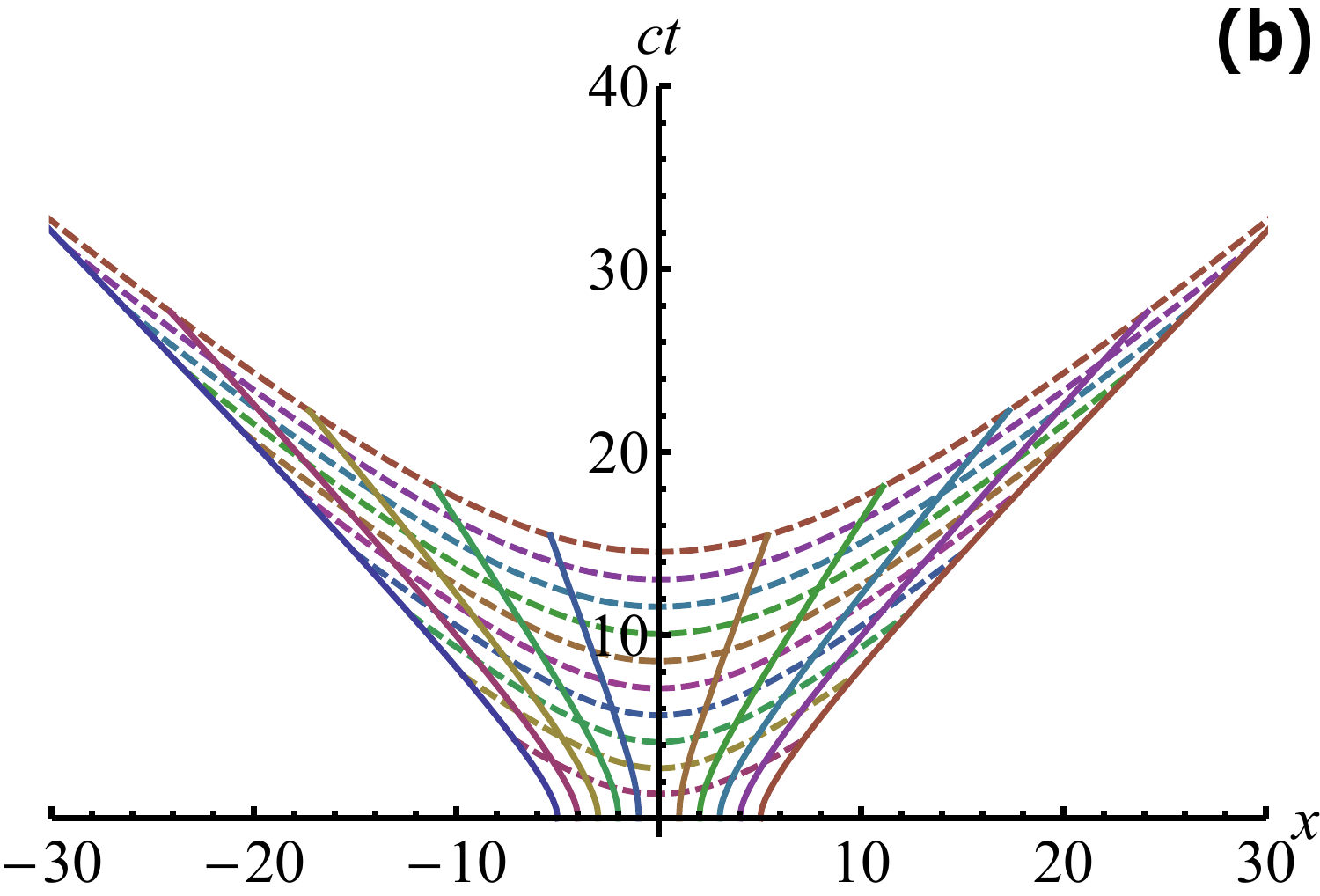}
\includegraphics[width=0.45\textwidth]{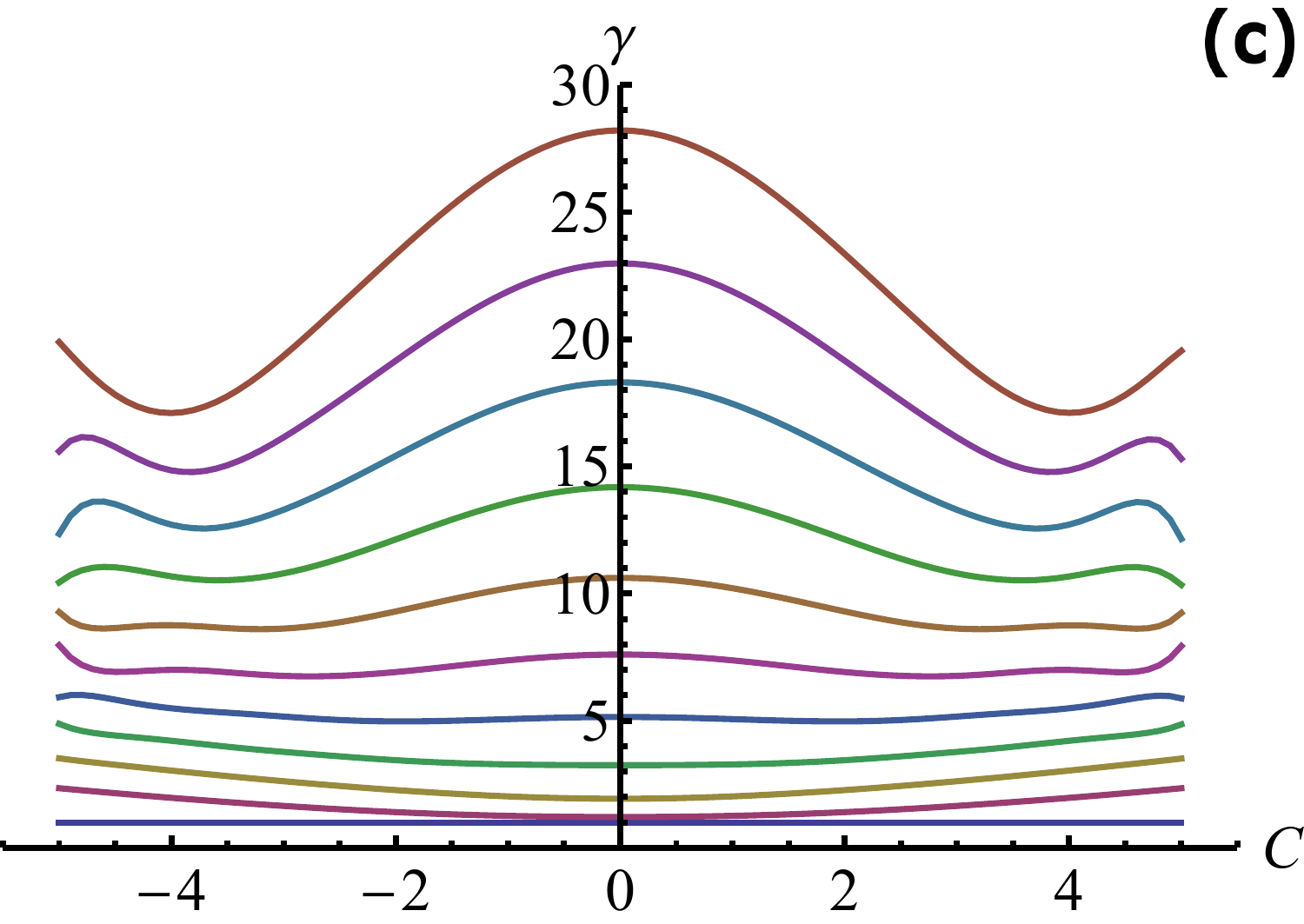}\hfill
\includegraphics[width=0.45\textwidth]{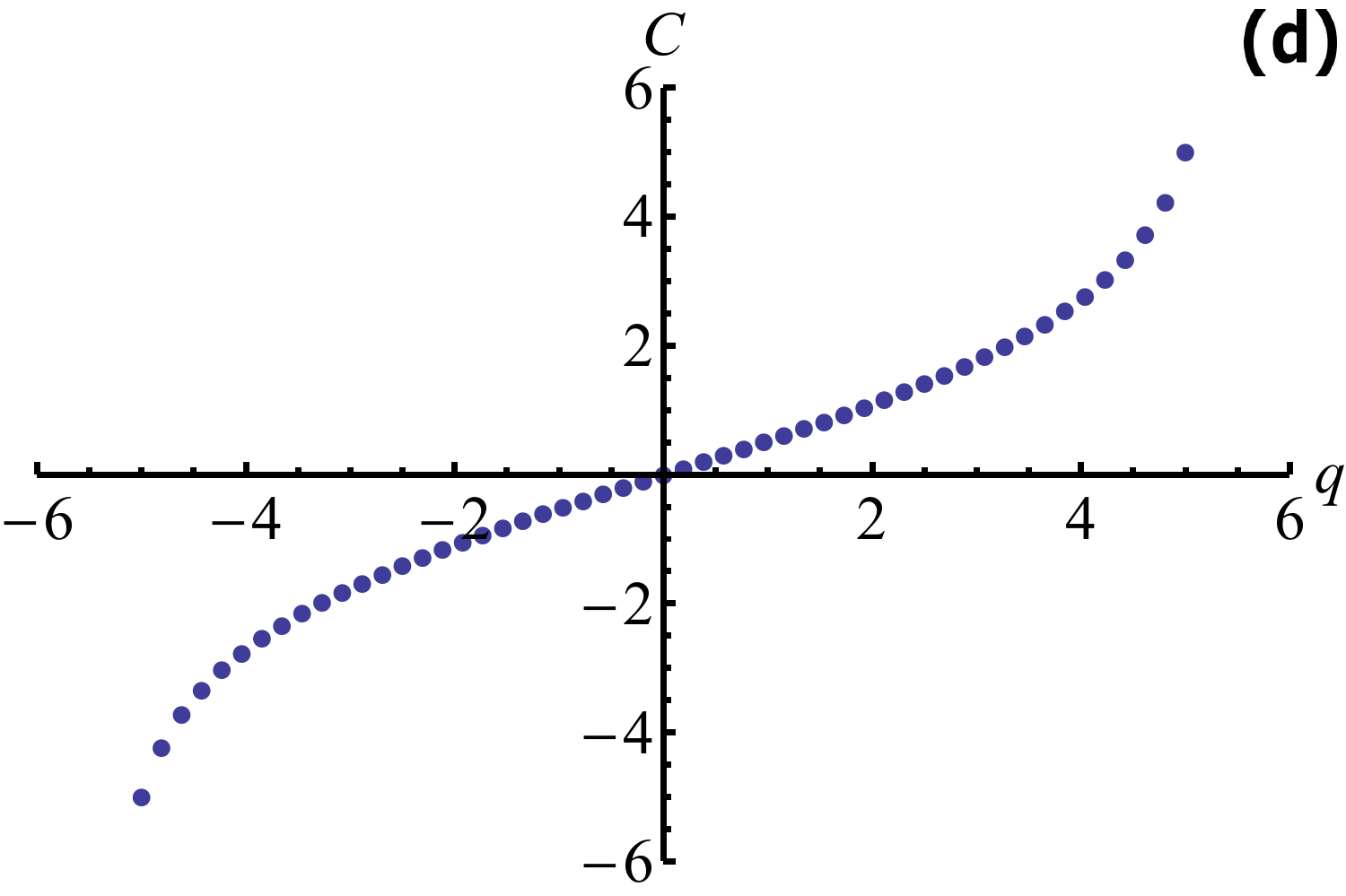}
\includegraphics[width=0.45\textwidth]{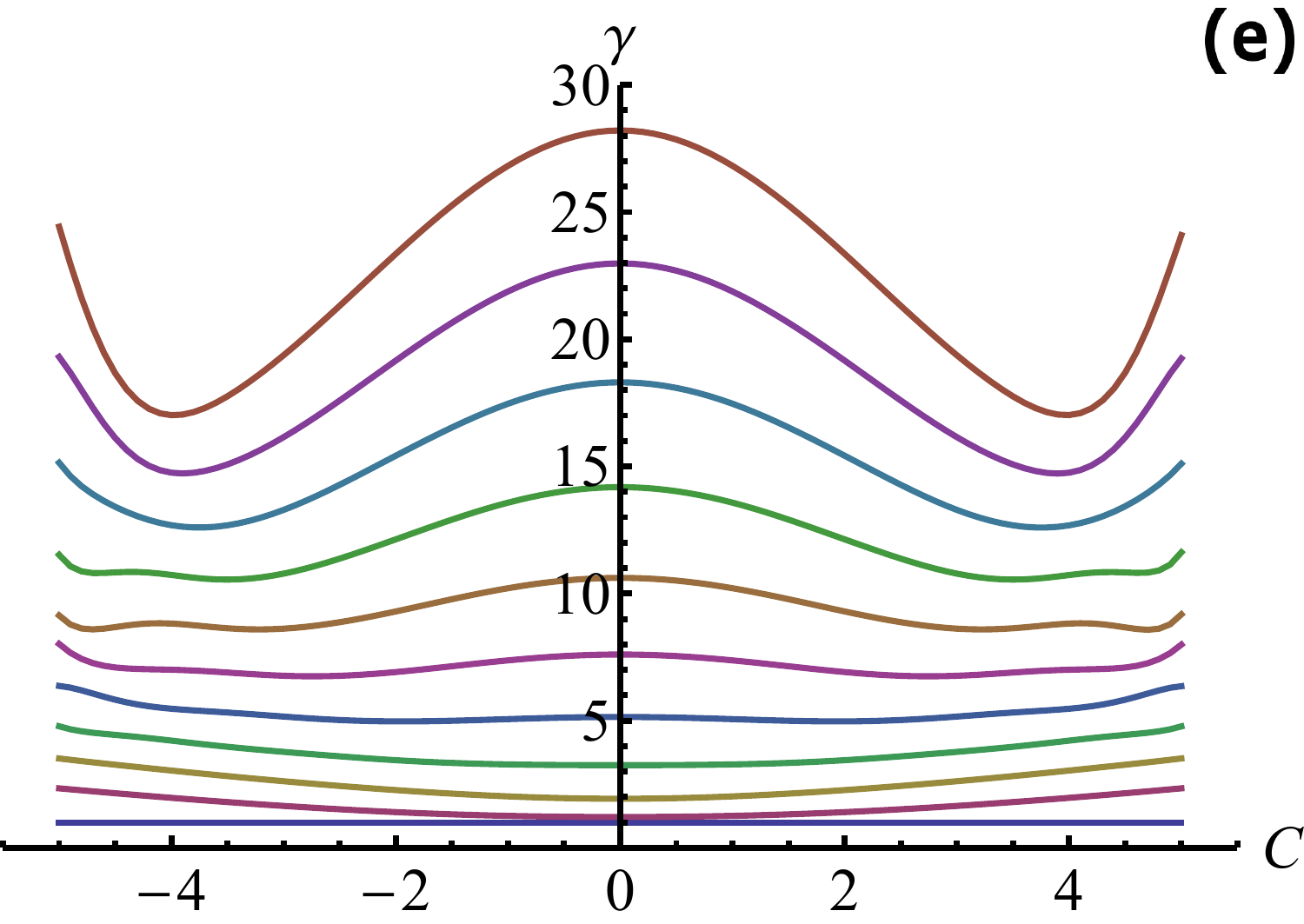}\hfill
\includegraphics[width=0.45\textwidth]{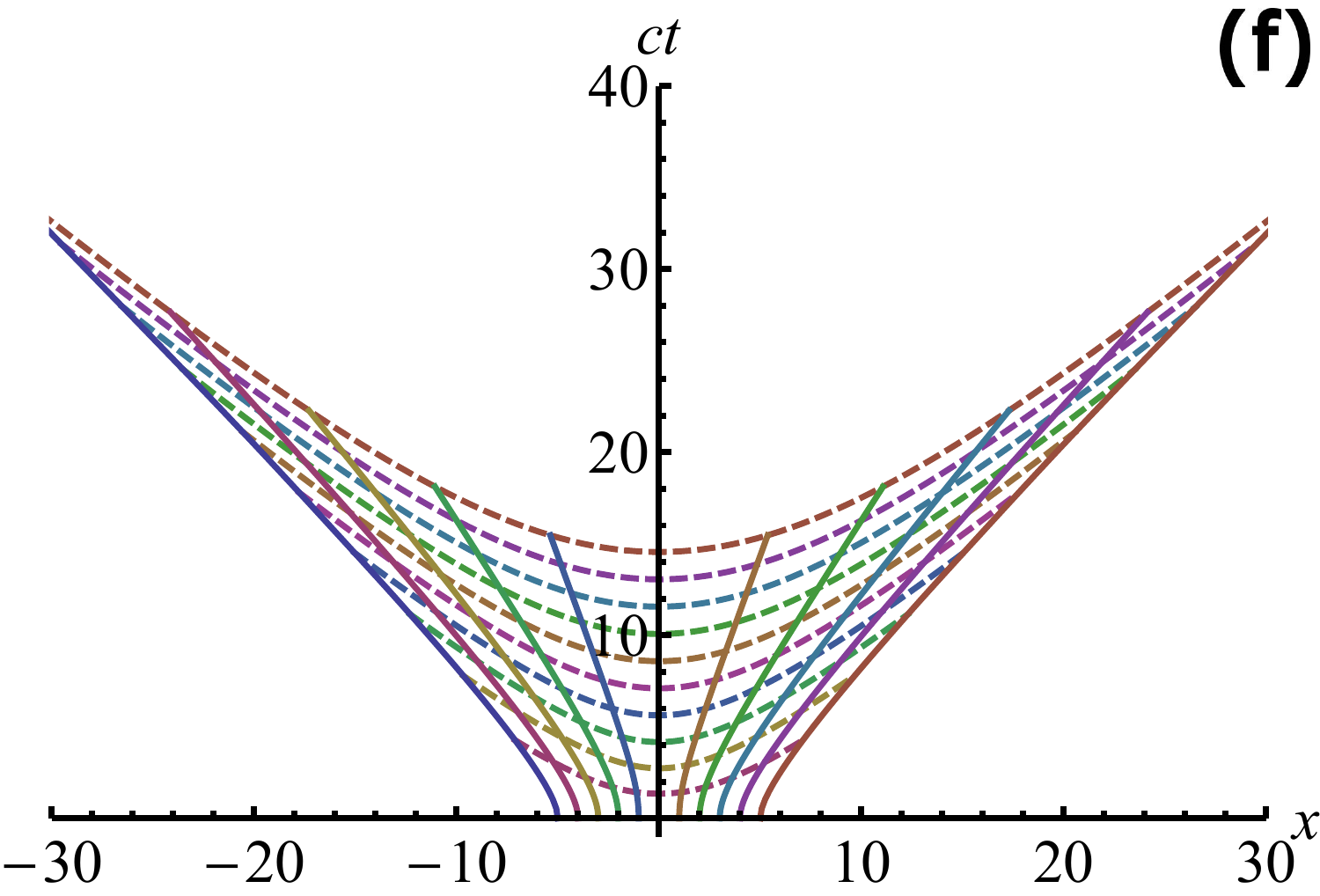}
\caption{Calculations with nonuniform $C$ grids: $N_g$ = 23 [(a) and (b)];  
$N_g$ = 53 [(c) and (d)]; $N_g$ = 93 [(e) and (f)].  Numerical $\gamma(\mathcal{T},C)$ curves
[(a), (c), and (e), as per Fig.~\ref{fig:SI-1}]  exhibit obvious discrepancies in the boundary regions,
but are very accurate in the interior.  Quantum trajectories [(b) and (f)] are nearly indistinguishable.
The $N_g=53$  grid is presented in (d).}
\label{fig:NUG}
\end{figure}

\begin{figure}
\includegraphics[width=0.45\textwidth]{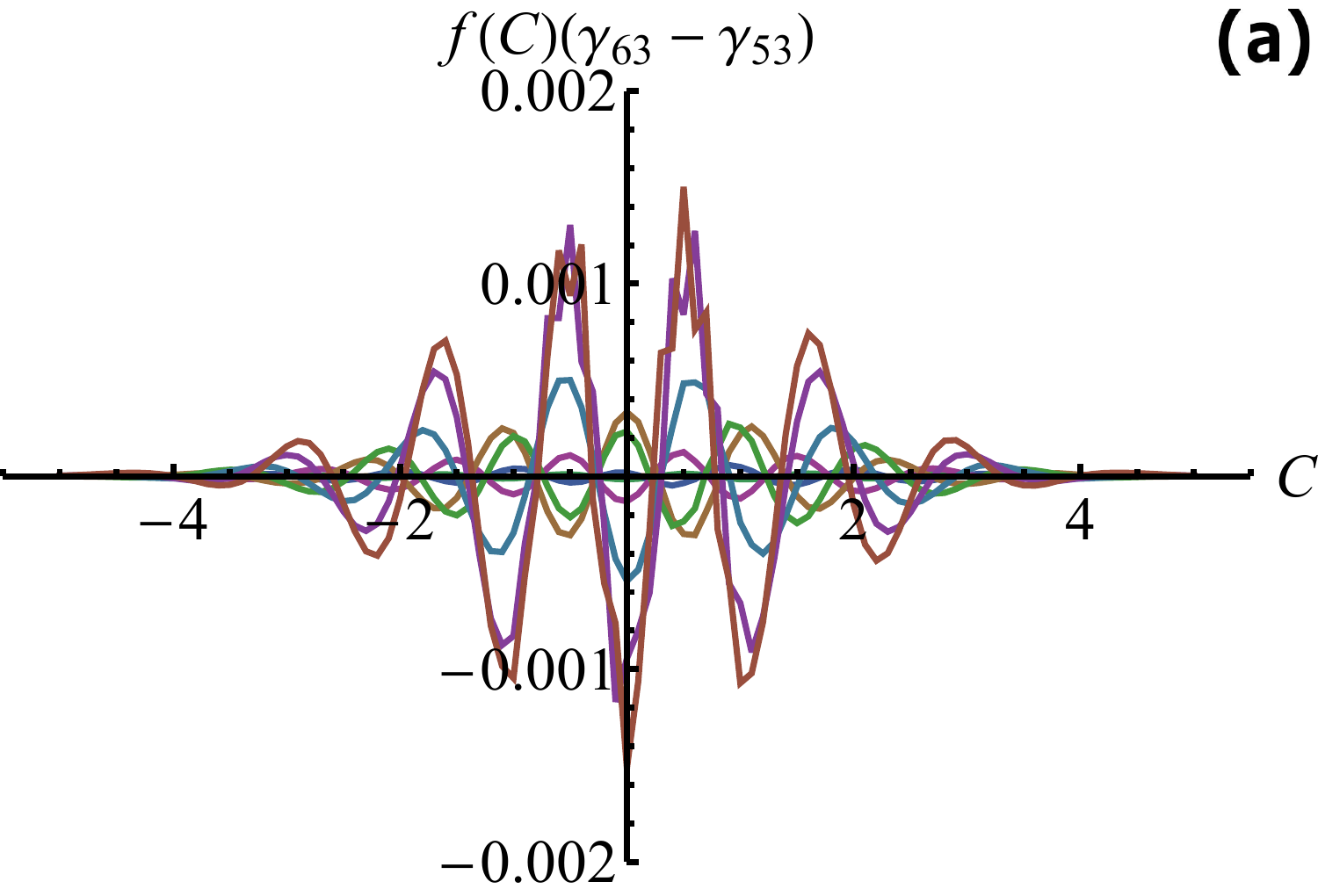}\hfill
\includegraphics[width=0.45\textwidth]{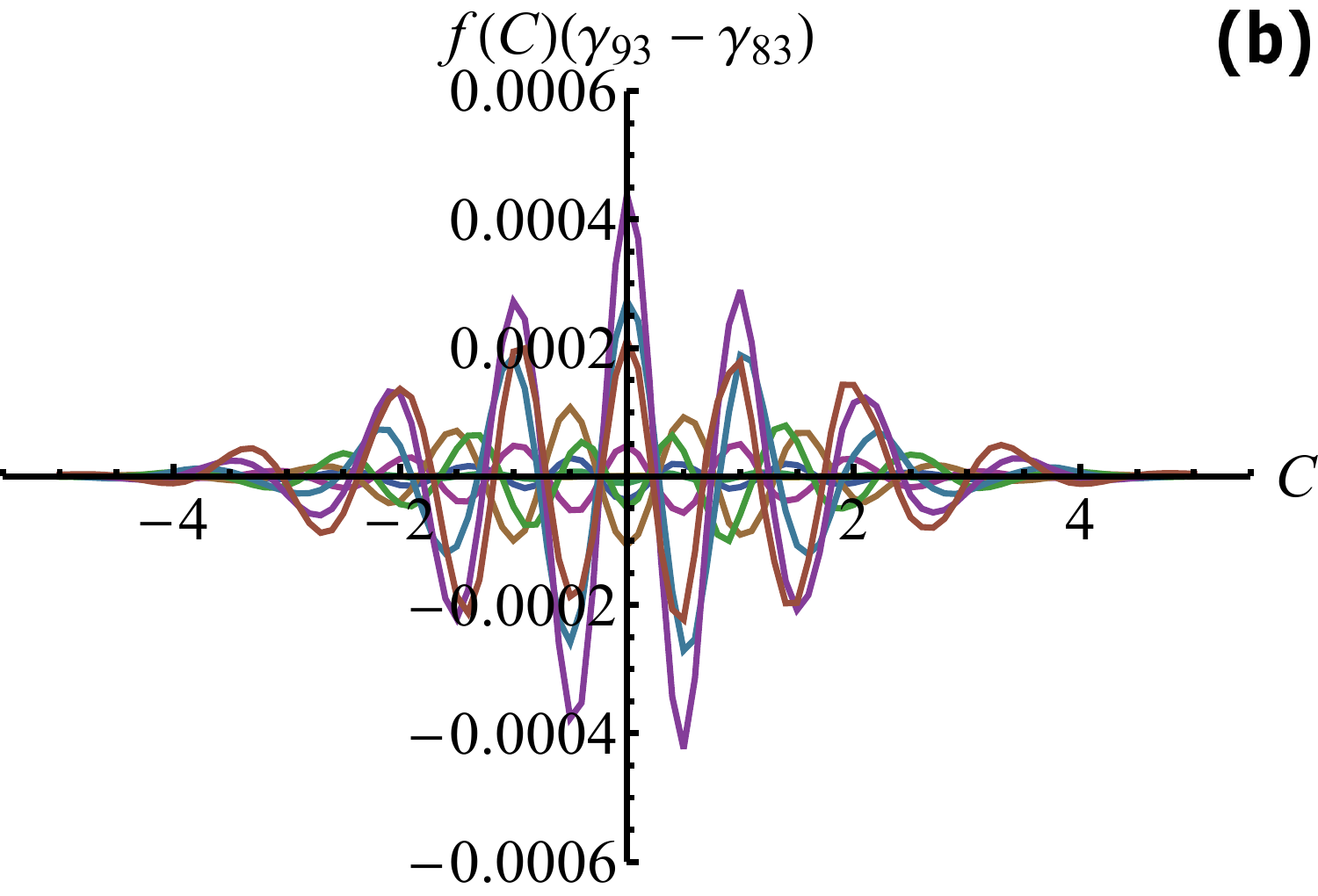}
\caption{Probability-weighted convergence errors for nonuniform $C$ grid calculations  of $\gamma(\mathcal{T},C)$,
as per Fig.~\ref{fig:SI-1}: (a) for $N_g=53$; (b) for $N_g=83$.
In each case, the error is defined relative to a slightly larger calculation, as $f(C)(\gamma_{N_g+10} - \gamma_{N_g})$.  
Numerical errors are substantially smaller for the larger grid, but in both cases the largest probability-weighted errors are in the interior.} 
\label{fig:NUG-6}
\end{figure}

For the above $c=1.5$ example, the choice
$(q_{\rm{max}} = 5, C_{\rm{max}} = 5, \beta= 0.19)$ enables stable, accurate calculations to be performed up to
$N_g=93$ grid points, as depicted in Fig.~\ref{fig:NUG}. 
Although numerical discrepancies in the boundary regions of the Fig.~\ref{fig:NUG} curves are still evident, we emphasize
that the wavepacket as a whole is \emph{extremely} well-converged.  This is evident in Fig.~\ref{fig:NUG-6}, which depicts the
\emph{probability-weighted} errors as a function of $C$ and $\mathcal T$,  for two different nonuniform grid calculations,
of sizes $N_g=53$ and $N_g=83$, respectively.  The larger grid size of the latter leads to substantial improvements in 
accuracy in the interior region. This is where the largest errors are to be found, in the probabilistic sense, although 
they are still quite small.  

\section{Probability conservation and Lorentz-boosted inertial frames \label{sec:LBF}}

In Secs.~\ref{sec:GWP-1d} and~\ref{sec:NUG}, we considered only free-particle Gaussian wavepackets which, 
at the initial time $\mathcal{T}=0$, comprise a stationary coherent state  (i.e., $\partial x /\partial \mathcal{T}|_{\mathcal{T}=0} =0$) 
that is centered at the origin [i.e., $x_0(C=0)=0$].  More generally, we wish to consider 
\emph{non}stationary wavepackets, for which the wavepacket center moves along an arbitrary (time-like) straight line.
The most general possible relativistic free-particle Gaussian wavepacket solutions may be obtained from the
stationary solutions through an arbitrary combination of: (a) translation in $t$; (b) translation in $x$; 
(c) Lorentz boost. Whereas (a) and (b) are trivial and will not be considered further, (c) serves as 
an explicit check on the Lorentz-invariance of the dynamical equations (despite the different PDE
orders in $\mathcal{T}$ and $C$), as well as the inertial coordinate 
probability conservation laws, as discussed in Sec.~\ref{sec:DE-3d}.

Let $S$ denote the inertial frame of reference $(ct,x)$, in which the free-particle Gaussian wavepacket
adopts the ``stationary'' form, in the sense described above.  Let $S'_{[v]}$ denote a new inertial frame, $(ct',x')$, 
obtained from $S$ via application of the Lorentz boost corresponding to the positive velocity $v>0$. 
In this manner, we obtain a new set of solutions, $\{x'^\alpha_{[v]}(X^\mu)\}$, each of which describes a 
quantum particle moving toward the negative $x'$-direction with ensemble-averaged velocity $-v$. 
For each boosted trajectory ensemble solution $x'^\alpha_{[v]}(X^\mu)$,  the  corresponding flux 
``four''-vector $j'^\alpha (x')$ can be easily obtained in the boosted frame $S'_{[v]}$.

\begin{figure}
\includegraphics[width=0.45\textwidth]{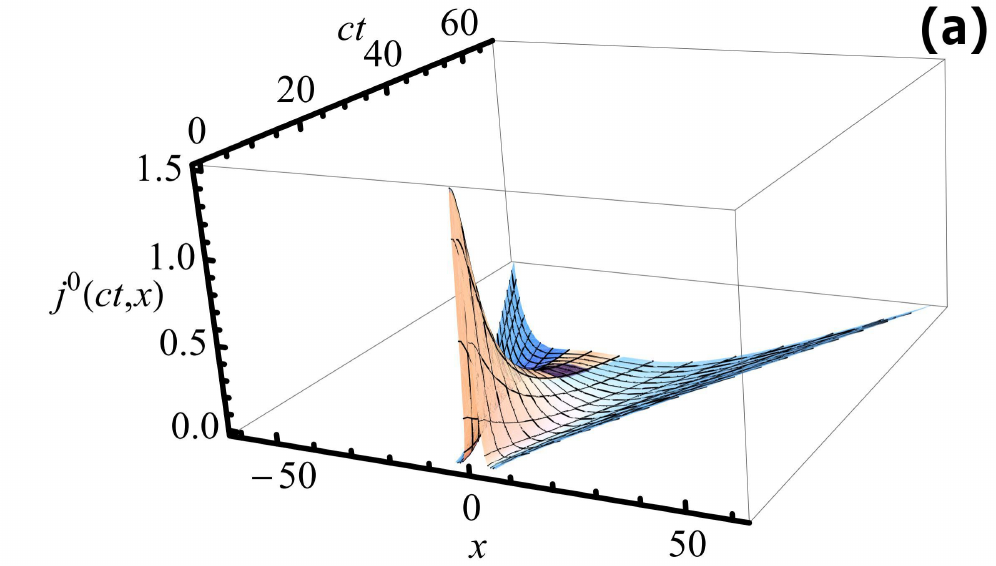}
\includegraphics[width=0.45\textwidth]{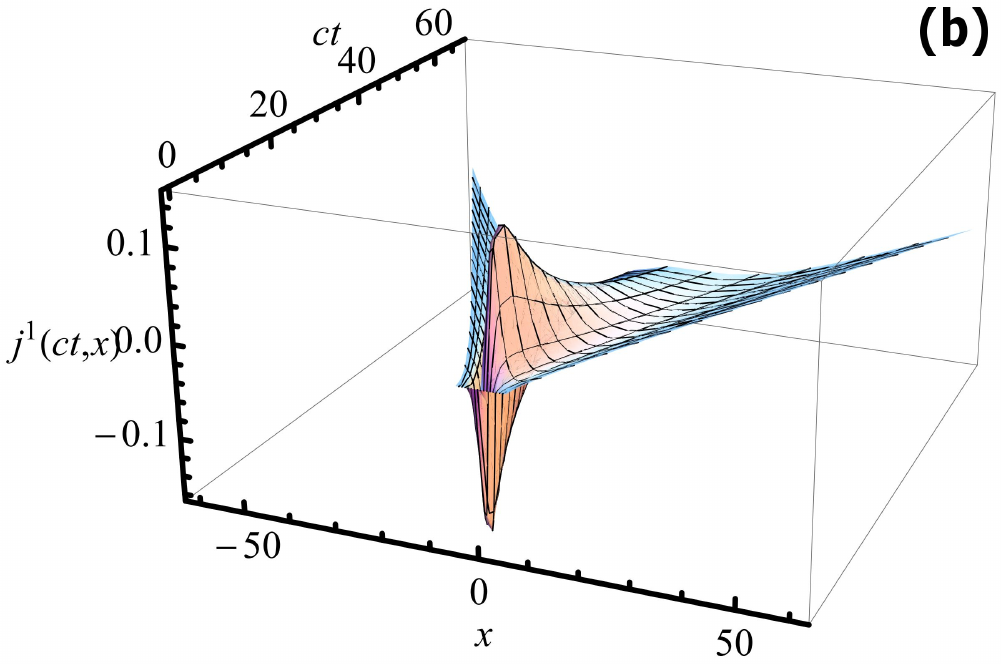}
\includegraphics[width=0.45\textwidth]{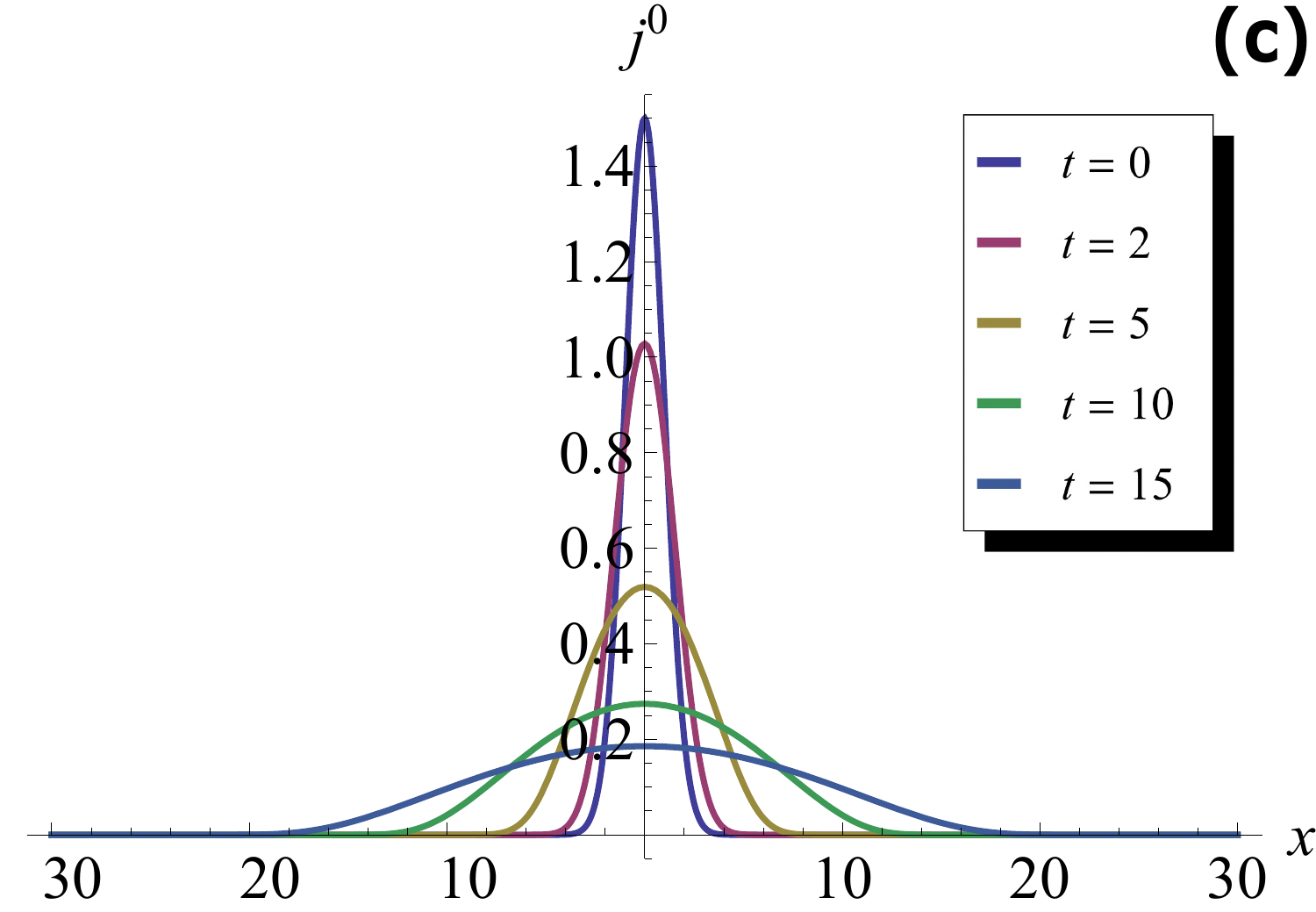} \hfill
\includegraphics[width=0.45\textwidth]{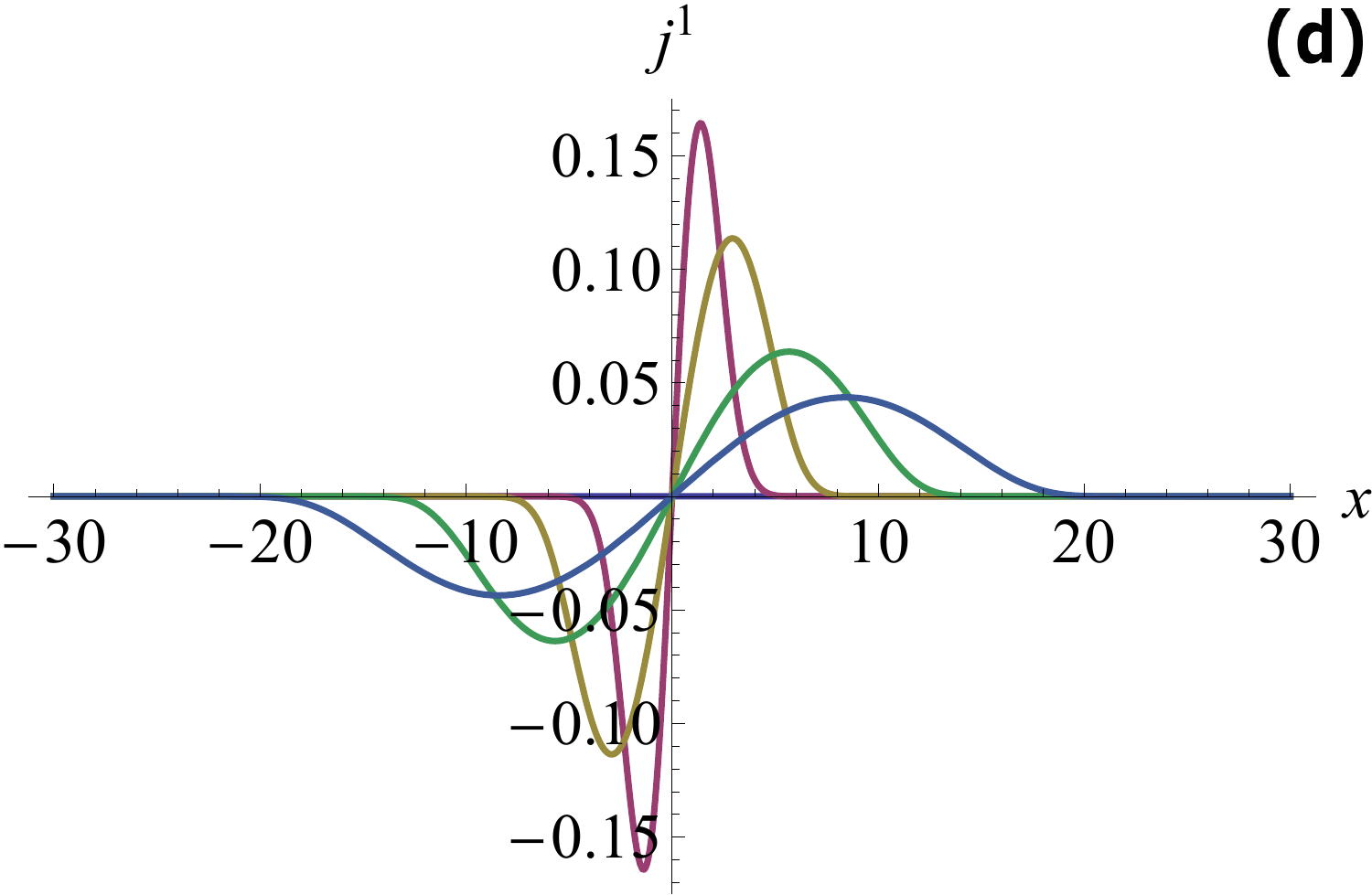}
\caption{Flux four-vector for the relativistic free-particle Gaussian wavepacket for $c=1.5$:
(a) three-dimensional plot of $j^0 (ct,x)$;  (b) three-dimensional plot of $j^1 (ct,x)$;  
(c) time slice plots of $j^0 (ct,x)$;  (d) time slice plots of $j^1 (ct,x)$. The time slices
are for  $t=\{0,2,5,10,15\}$.}
\label{fig:j_0_1}
\end{figure}

It is of interest to verify the conservation of probability [Eq.~(\ref{Pcons})] and covariant continuity
relation [Eq.~(\ref{continuity})] of Sec.~\ref{sec:DE-3d} numerically, in both stationary and boosted 
inertial frames.  We do so for the stationary frame as follows. 
The stationary inertial flux four-vector is expressed as \cite{Poirier:2012gz},
\ba
j^\alpha (x) = \frac{f(C)}{\sqrt{\gamma}} \frac{d x^\alpha}{d \tau} = \frac{e^{-a C^2}}{\sqrt{\gamma}} \frac{d x^\alpha}{d \tau}. \label{eq:j_alpha_1}
\ea
We obtain the probability density and flux, respectively, as follows:
\ba
j^0 (ct, x) = c \frac{f(C)}{\sqrt{\gamma}} \frac{dt}{d \tau}\qquad ; \qquad
j^1 (ct, x) =  \frac{ f(C) }{ \sqrt{ \gamma } } \frac{d x}{d \tau}, \label{eq:j-expression}
\ea
where
\ba
\frac{dt}{d \tau} =  \frac{1}{\sqrt{1-\frac{1}{c^2} \left(  \frac{dx }{dt }\right)^2  }} \qquad ; \qquad
\frac{dx}{d \tau} = \frac{\frac{dx }{dt }}{\sqrt{1-\frac{1}{c^2} \left(  \frac{dx }{dt }\right)^2  }}. \label{eq:dtau}
\ea
To obtain $j^\alpha (ct,x)$ numerically, we start with the numerical quantum trajectory ensemble solution, 
$x^\alpha(X^\mu) = \bigl( ct(\mathcal{T},C), x(\mathcal{T},C) \bigr)$. From the latter we obtain $dx/dt$ everywhere via
\ba
\frac{dx}{dt} = \left.\frac{\partial x}{\partial \mathcal{T}}\right|_C \left( \left. \frac{\partial t}{\partial \mathcal{T}} \right|_C \right)^{-1}. \label{eq:dx_dt-1}
\ea
Substituting Eq.~(\ref{eq:dx_dt-1}) into Eq.~(\ref{eq:dtau}), and again into Eq.~(\ref{eq:j-expression}), we obtain
numerical results for $j^0(ct, x)$ and $j^1 (ct,x)$.  These are presented in Fig.~\ref{fig:j_0_1}, for the $c=1.5$ example.

In Fig.~\ref{fig:j_0_1}(a) and (c), we observe that $j^0 \left(c t,x\right) > 0$ for all $(ct, x)$---i.e., the probability density remains
positive always. We also observe substantial broadening of the wavepacket over time. However, the initial Gaussian
shape is also substantially distorted over time, with the fringe trajectories ``bunching up'' as they approach the speed
of light (Sec.~\ref{sec:SI}). As for the flux, $j^1 \left( ct,x\right)$, it is always odd in $x$, being positive for $x>0$
and negative for $x<0$ (for $t \ge 0$). This also reflects wavepacket broadening---i.e., trajectories to the right of $x=0$ spread 
to the right, whereas those to the left spread to the left. Although the trajectory speeds increase monotonically with 
increasing $|C|$, the flux reaches a maximum and then decays, owing to the probability weighting. Finally, according
to the $(1+1)d$ Eq.~(\ref{Pcons}), the areas under each of the $j^0(x)$ curves in Fig.~\ref{fig:j_0_1}(c) should 
be constant, if probability is conserved. Numerical integration for each of the sixteen $t$ values, $t=\{0, 1, 2, ..., 15\}$,
 results in a 
near-constant value of $3.7615$, with an RMS error of only $0.0011$.
 
In the Lorentz-boosted frame  $S'_{[v]}$, we denote the flux four-vector as $j'^\alpha (ct', x')$.
Being a true Lorentz-invariant vector, this quantity must  transform as does $x^\alpha$---i.e., as 
\ba
x'^\alpha = \tensor{\Lambda}{^\alpha_\delta} \,x^\delta \qquad ; \qquad
j'^{\alpha}  (ct', x') =  \tensor{\Lambda}{^\alpha_\delta} \,  j^\delta  (ct, x),
 \label{eq:LB_transform}
\ea
where the rank-2 boost tensor $\tensor{\Lambda}{^\alpha_\delta}$ can be expressed in terms of the
parameter $\beta= v/c$ as 
\begin{eqnarray}
\tensor{\Lambda}{^\alpha_\delta} = \left( \begin{array}{cc}
\frac{1}{\sqrt{1-\beta^2}} & \frac{-\beta}{\sqrt{1-\beta^2}} \\
\frac{-\beta}{\sqrt{1-\beta^2}} & \frac{1}{\sqrt{1-\beta^2}} \\
\end{array} \right),  \label{eq:LB_matrix_1}
\end{eqnarray}
in the usual manner.  

\begin{figure}
\includegraphics[width=0.45\textwidth]{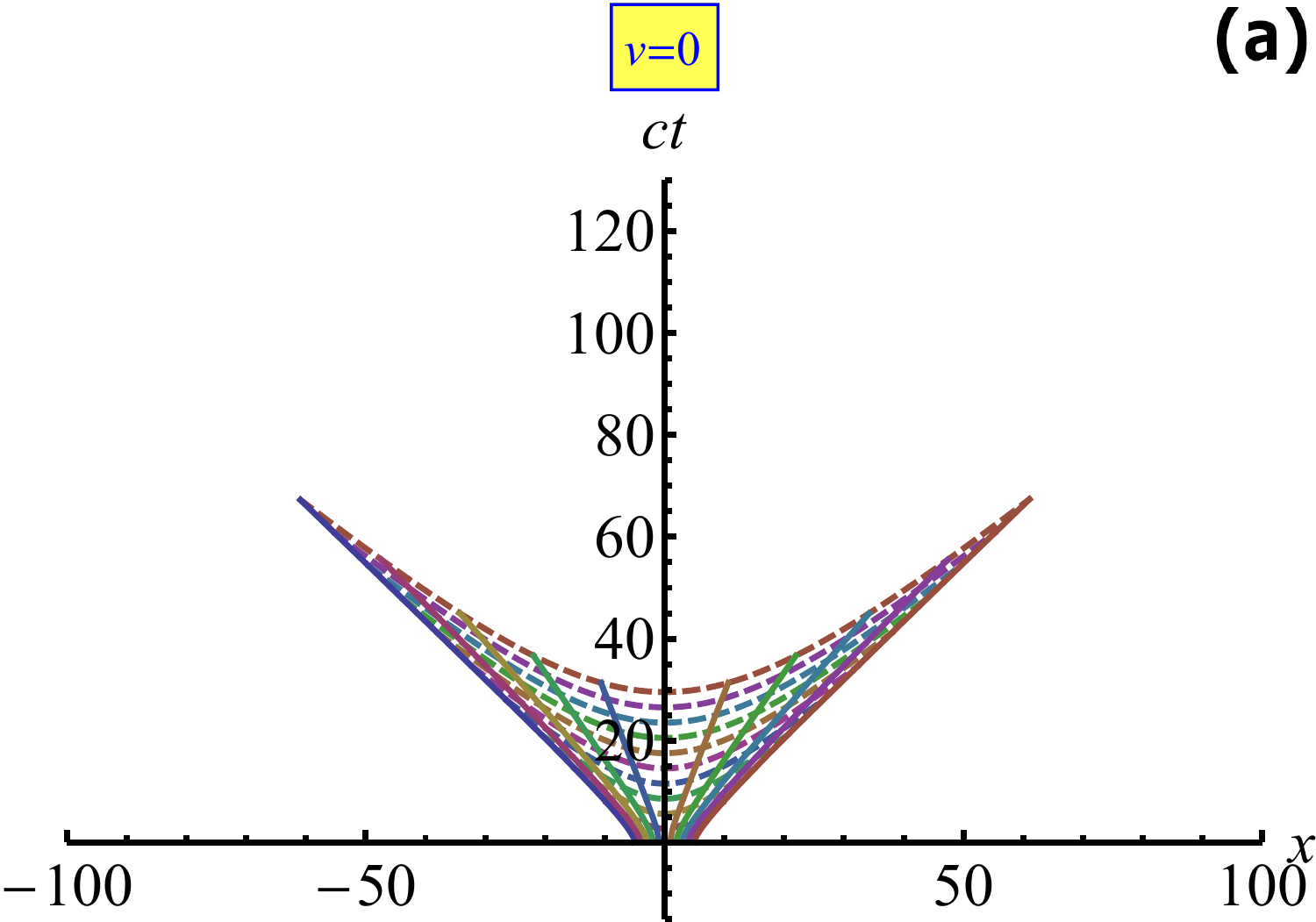}
\includegraphics[width=0.45\textwidth]{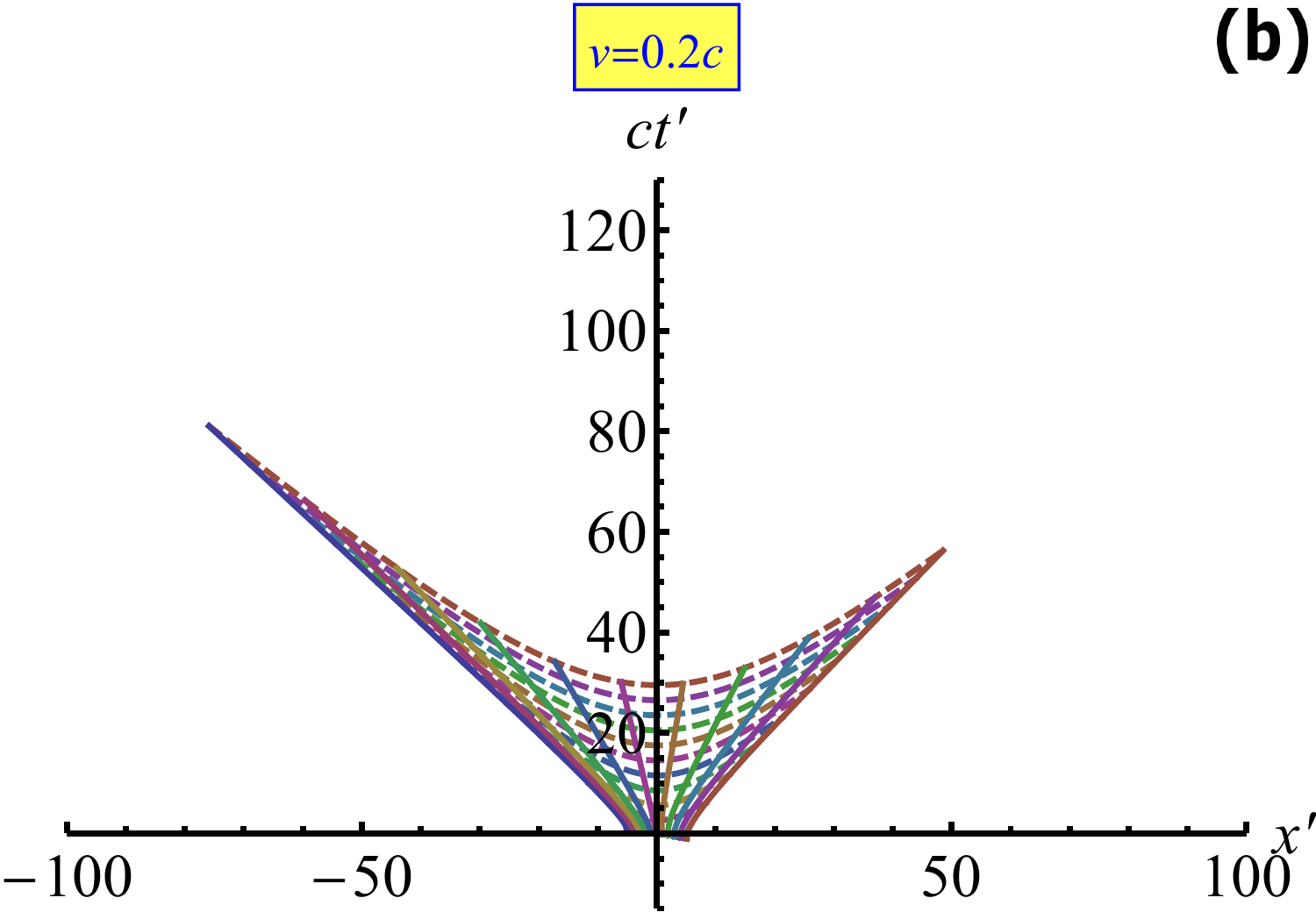}
\includegraphics[width=0.45\textwidth]{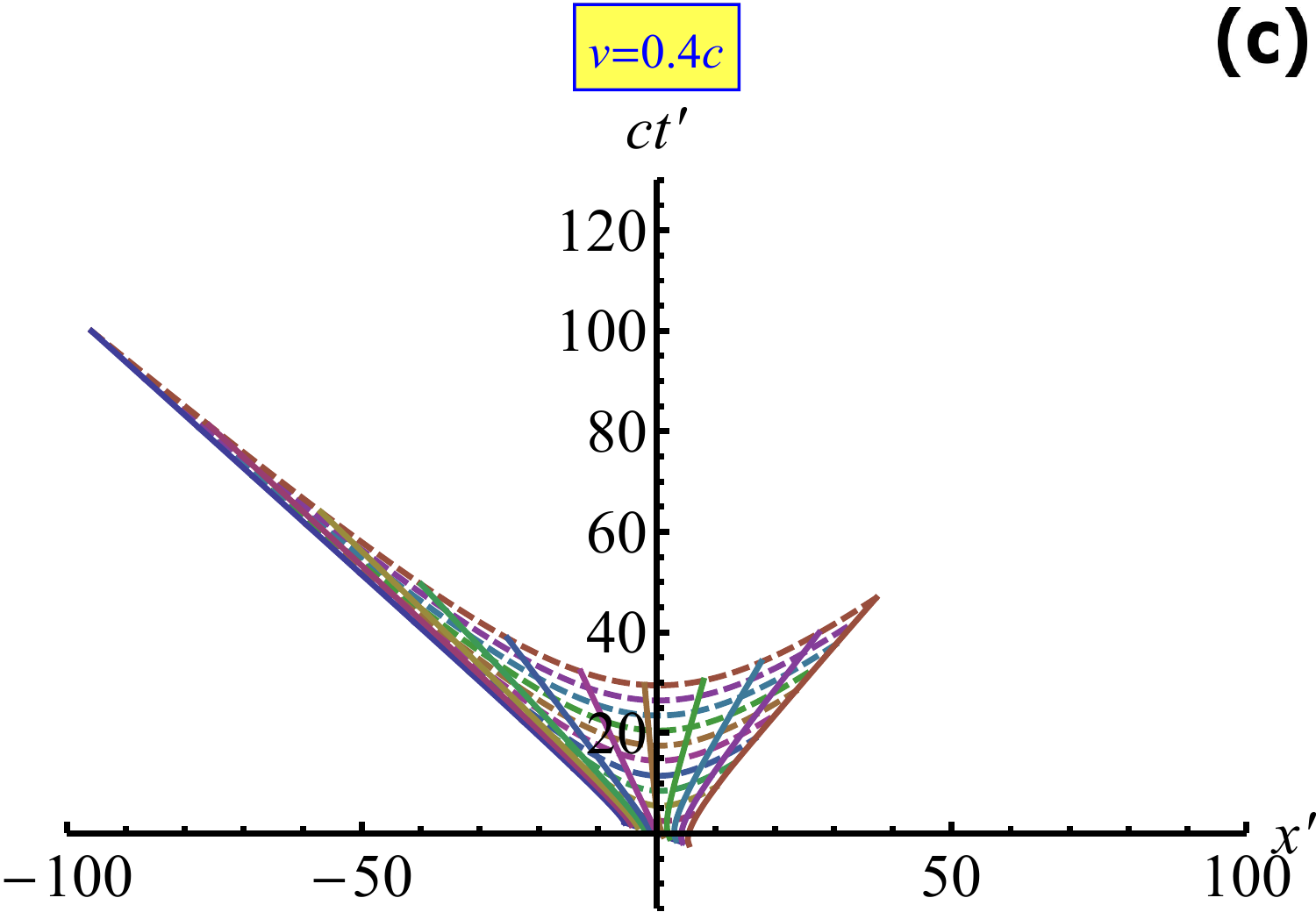}\hfill
\includegraphics[width=0.45\textwidth]{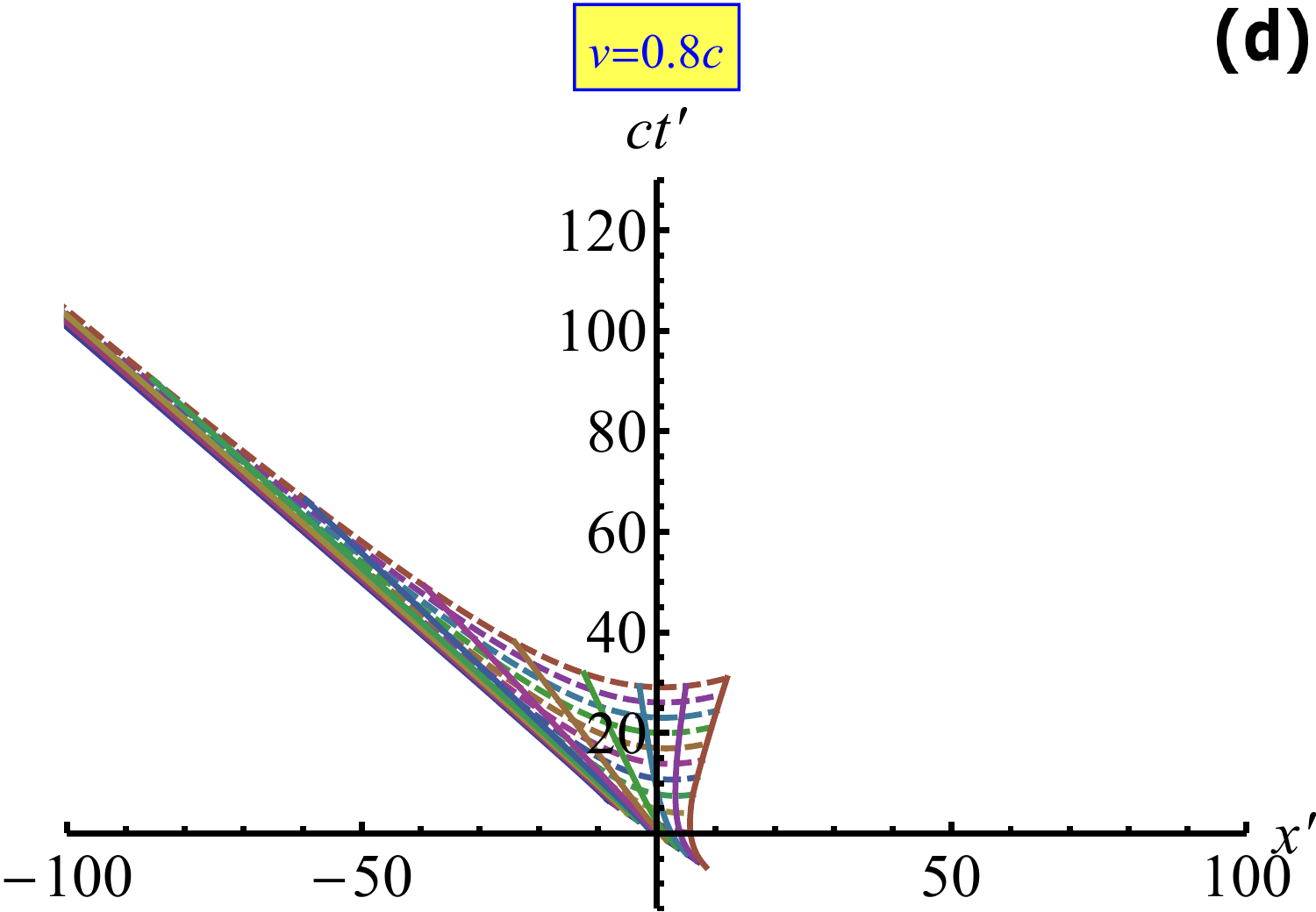}
\caption{Quantum trajectories (solid curves) and simultaneity submanifolds (dashed curves) 
 for the relativistic free-particle Gaussian wavepacket for $c=1.5$, 
in various stationary and Lorentz-boosted frames: 
(a) $v=0.0c$; (b) $v=0.2c$, (c) $v=0.4c$; (d) $v=0.8c$.}
\label{fig:traj_prime}
\end{figure}

In the Lorentz-boosted frame $S'_{[v]}$, a \emph{different} probability conservation
law than Eq.~(\ref{Pcons}) must hold, namely, 
\ba
        \int j'^{0}(ct', x')\, d x'  = {\rm const}.
        \label{Pcons-prime}
\ea
Once again, we will verify numerically that the Eq.~(\ref{Pcons-prime}) probability conservation law
is indeed satisfied.  First, however, we obtain the trajectory ensemble solution, 
$x'^\alpha(X^\mu)$, by Lorentz-boosting $x^\alpha(X^\mu)$ (dropping the `$[v]$' subscript).
Then, we apply a second Lorentz boost to the dependent variables $j^\alpha = (j^{0}, j^{1})$, to obtain 
the desired functions $j'^{0}(ct', x')$ and $j'^{1}(ct', x')$.

\begin{figure}
\includegraphics[width=.45\textwidth]{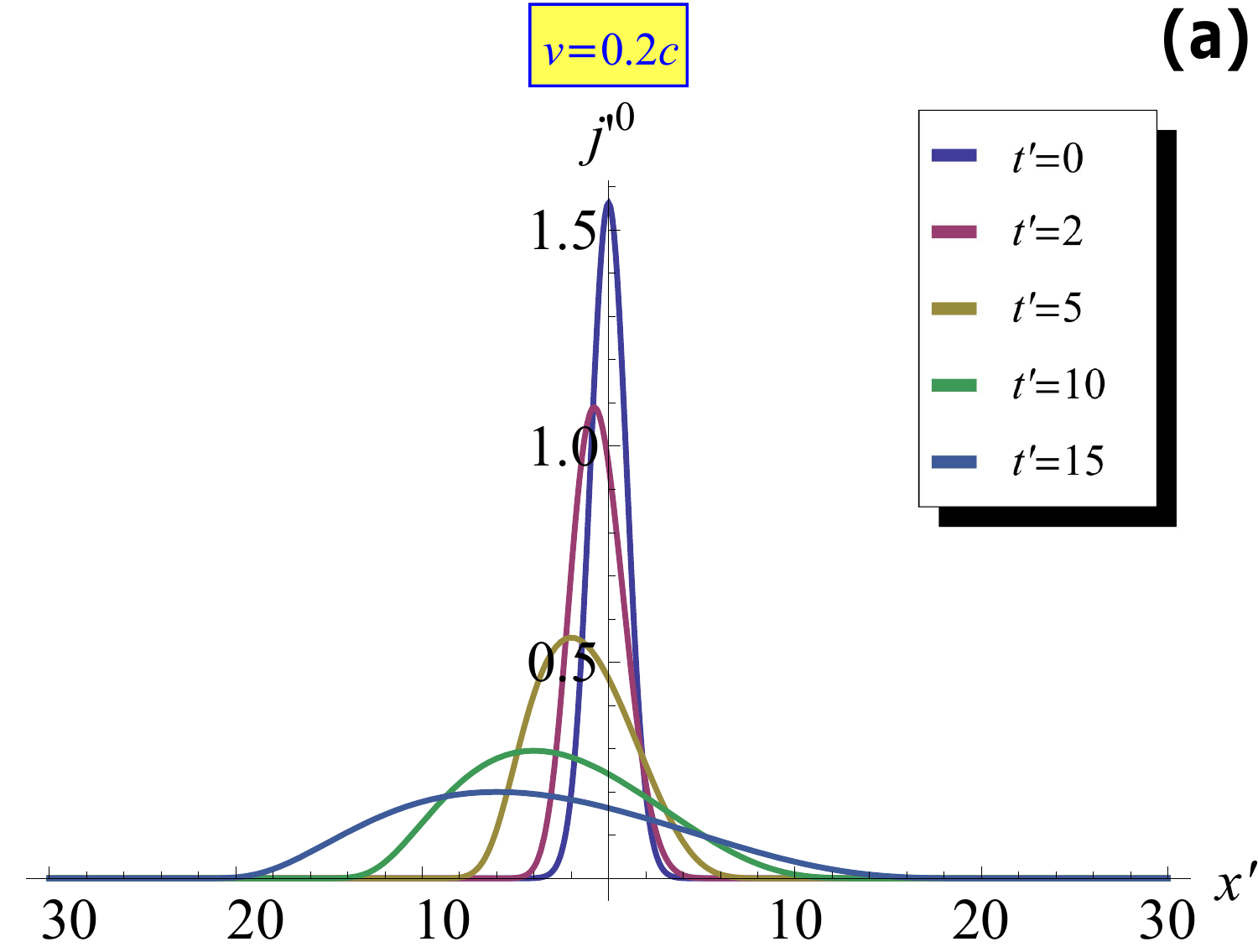}
\includegraphics[width=.45\textwidth]{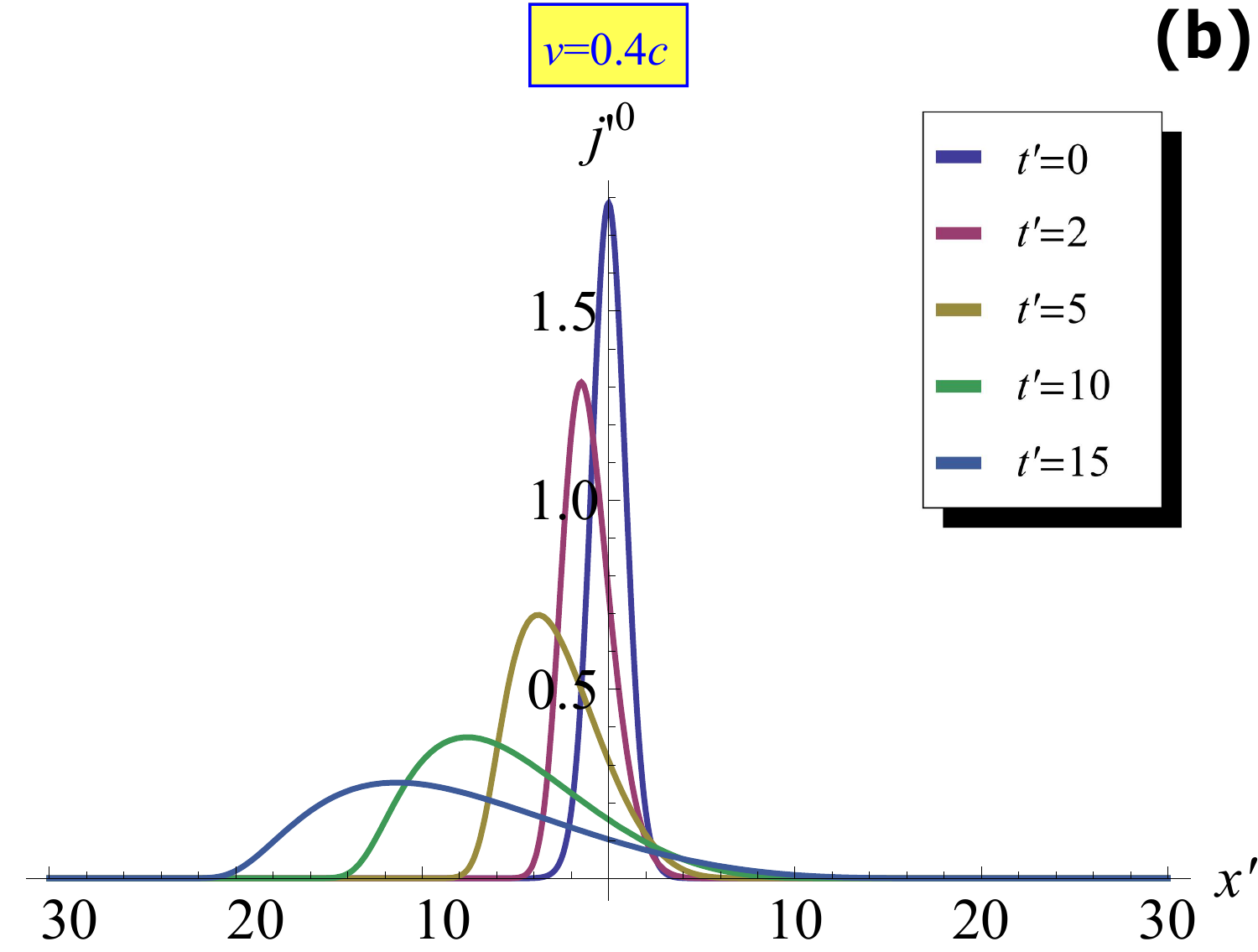}
\includegraphics[width=.45\textwidth]{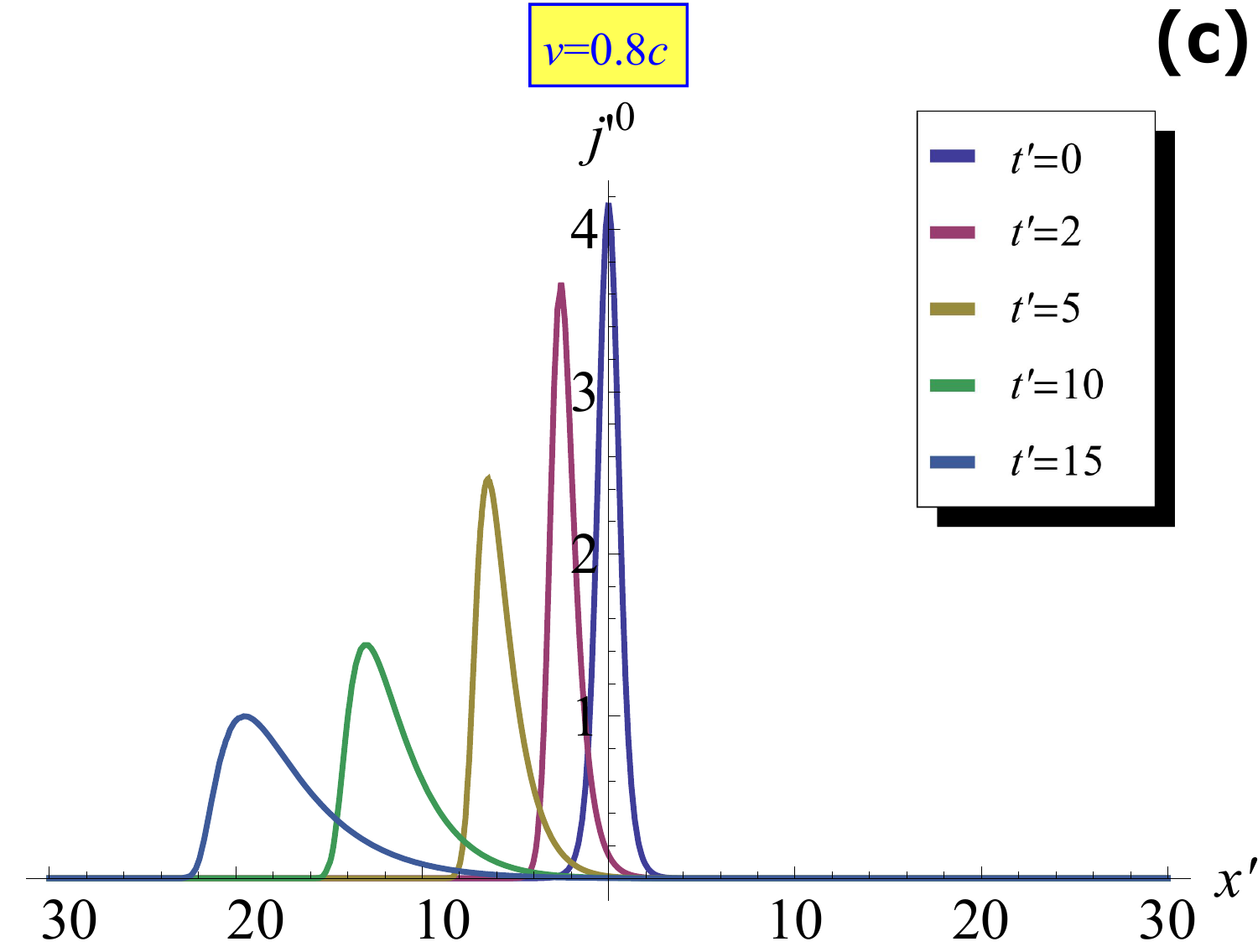}\hspace{3pc}%
\begin{minipage}[b]{18pc}\caption{\label{fig:j0_prime}
Time slice plots of the probability density $j'^0(ct',x')$, for the relativistic free-particle 
Gaussian wavepacket for $c=1.5$, in various Lorentz-boosted frames: 
(a) $v=0.2c$, (b) $v=0.4c$;  (c) $v=0.8c$. The time slices are for  $t'=\{0,2,5,10,15\}$.}
\end{minipage}
\end{figure}

In Fig.~\ref{fig:traj_prime}, we depict the quantum trajectories and simultaneity submanifolds in the 
stationary frame $S$,  and in several boosted frames  $S'_{[v]}$ for $v=0.2c$, $v=0.4c$ and $v=0.8c$. 
The boosted quantum trajectories do indeed move generally to the left, increasingly so for increasing $v$. 
The trajectory ensemble is no longer symmetric in $x'$, but it is symmetric under $(ct',x') \rightarrow (-ct',-x')$, 
as would be expected from the nonrelativistic case. 
Figure~\ref{fig:j0_prime} depicts $j'^0(ct', x')$, plotted as a function of $x'$ for various
time slices $t'$, for $v=0.2c$, $v=0.4c$ and $v=0.8c$. In all cases, the motion of the wavepacket 
to the left over time is evident, as is wavepacket broadening.  We also observe
a  ``bunching'' distortion of the Gaussian form in the wavepacket fringes, similar  to that evident 
in Fig.~\ref{fig:j_0_1}(c) for the stationary case.  Here, however,  for all $t' \ne 0$, the wavepacket is not
even symmetric about its center---unlike both the travelling nonrelativistic Gaussian
wavepacket, and the stationary relativistic example discussed previously.  The reason is clear: due to
the velocity bias in the ensemble, trajectories to the left of the wavepacket center experience much more
pronounced bunching than those to the right---which may not even bunch at all, if the region 
of significant probability is always moving to the left over the time scale of interest. 

Finally, we consider conservation of probability.  First, note that  according to 
Figure~\ref{fig:j0_prime}, $j'^0 \left(c t',x'\right) \ge 0$ for all $(ct', x')$---thus 
justifying our interpretation of this quantity as a probability density.  Also, once again, 
for Eq.~(\ref{Pcons-prime}) to be satisfied, the areas under each of the curves in Fig.~\ref{fig:j0_prime}
must be the same, for a given $v$ value.  Via numerical integration, this is indeed found to be the case.
Specifically:  for $v=0.2c$, the integration for each curve yields $3.8369$, with an RMS error of  $0.0008$; 
                        for $v=0.4c$, the integration for each curve yields $4.1029$, with an RMS error of  $0.0009$; 
                        for $v=0.8c$, the integration for each curve yields $6.2704$, with an RMS error of  $0.0097$. 
Thus, we conclude that conservation of probability as per Eq.~(\ref{Pcons-prime}) holds
in all Lorentz-boosted frames, $S'_{[v]}$, and therefore for all relativistic free-particle Gaussian wavepackets.

\section{Summary and Conclusions}

In this paper, we analyze the dynamical equations for a single, spin-zero relativistic 
quantum particle, based on a recently-derived trajectory-based formulation \cite{Poirier:2012gz}. The PDE 
[Eq.~(\ref{eq:rel_PDE-3d})] governing the trajectory ensemble solution $x^\alpha(X^\mu)$ is second
order in \emph{natural} time $\mathcal{T}$, and fourth-order in \emph{natural} space $\bf{C}$,
yet treats \emph{inertial} $t$ and $x$ on an equal, Lorentz-invariant
footing.  As a result, the main issues plaguing the traditional wave-based equations
of Klein-Gordon and Dirac (when viewed as single-particle theories) are resolved here.

This is especially true for wavepacket dynamics, which serves as the focus of this contribution. 
Note that even though the traditional wavefunction plays no role in the trajectory formulation,
one can nevertheless extract a wavepacket description from the latter---in terms of the 
inertial probability density quantity, $j^0$, which in turn can be obtained from the $x^\alpha(X^\mu)$
solution itself, together with the natural spatial scalar probability density $f(\bm{C})$.  Unlike 
wave-based relativistic quantum treatments, the wavepacket dynamics observed here for a free particle
are always simple, straightforward, and well-behaved.  In particular, $j^0$ is everywhere positive, 
localized, and interference-free, with an ensemble-averaged position that travels along a linear path 
through spacetime.  Zitterbewegung and other  complications---stemming ultimately from interference 
between positive- and negative-energy wave solutions---never arise here, because there are no 
negative-energy trajectory solutions. 

We have also examined the symmetry properties of the $(3+1)d$ and $(1+1)d$ dynamical equations,
and of their numerically computed Gaussian wavepacket solutions---with respect to parameter and 
dynamical variable rescaling, as well as Lorentz boosts.  The invariance with respect to these 
symmetry operations has been verified both analytically and numerically.  In particular, from 
scale invariance, we learn that narrowing the initial Gaussian width (by increasing $a$) 
is essentially equivalent to reducing the speed of light $c$ (keeping both $m$ and $\hbar$ fixed). 
This makes good physical sense;
even in the nonrelativistic context, narrower initial coherent wavepackets are known to 
disperse more rapidly than broader ones---corresponding in the present context to 
significant-probability trajectories that approach the speed of light more rapidly. In the extreme
ultrarelativistic limit $c \rightarrow 0$, one accordingly obtains a Dirac delta initial condition, 
whose corresponding singular solution has been obtained analytically \cite{Poirier:2012gz}, and may 
be relevant for cosmology.  In the opposite scaling limit, one approaches the standard 
nonrelativistic Gaussian wavepacket solution---except for the $|C|\rightarrow \infty$ ``fringe'' trajectories that carry 
almost no probability.    

In this work, only the stationary $x^\alpha(X^\mu)$ solutions were obtained explicitly numerically---although
the term ``stationary'' is a bit of a misnomer, since quantum forces lead to accelerating trajectories
that give rise to wavepacket spreading over time. To obtain completely general, moving  
Gaussian wavepacket solutions, Lorentz-boosting is applied to the stationary solutions.  In
the wave-based approach, Lorentz boosts in and of themselves can introduce undesirable 
interference effects \cite{thaller}.  This is not the case for the present trajectory-based
approach---e.g., the transformed $j'^0$ is just as well behaved as $j^0$, and $x^\alpha$ and 
$j^\alpha$ exhibit perfect Lorentz invariance. 

Numerical solution of the Eq.~(\ref{eq:PDE_two}) PDE is complicated by the fact that the 
boundary conditions are not known \emph{a priori}. The simple remedy proposed here, 
involving a nonuniform $C$ grid, enables higher grid densities and accuracies to be obtained for the
physically relevant trajectories that bear significant probability.  On the other hand,
experience with the nonrelativistic case strongly suggests that this approach will not be so effective
in the presence of external fields.  Thus, more robust strategies will have to be developed, going
forward.   

The present work and Ref.~\cite{Poirier:2012gz} serve to demonstrate that the trajectory-based formulation
is a very promising approach in the fixed-particle-number relativistic quantum context, worthy of 
further development. 
One very convenient feature of this approach is that it derives  from a trajectory-based Lagrangian. 
Thus, both the nonrelativistic and classical limits emerge straightforwardly. Moreover, the Euler-Lagrange 
and Noether procedures can be directly applied, to obtain the dynamical PDE and conservation
laws, respectively. All of these aspects  will be examined in future publications. Another,
quite obvious generalization will also be considered: just as the present approach constitutes
the trajectory-based analog of the Klein-Gordon equation, by incorporating spin into the theory,
we hope to derive the trajectory version of the Dirac equation. Finally, one is not restricted
to a trajectory-based approach. Recent work~\cite{Poirier:2012gz} suggests that a mathematically 
equivalent wave equation can also be developed, which may offer certain advantages, although
it would necessarily have to be nonlinear.

\ack{This work was supported in part by a grant from the Robert A. Welch Foundation 
(D-1523).  The authors also acknowledge useful discussions with Jeremy Schiff.}


\section*{References}

\bibliography{rel-wave}

\providecommand{\newblock}{}
\begin{thebibliography}{10}
\expandafter\ifx\csname url\endcsname\relax
  \def\url#1{{\tt #1}}\fi
\expandafter\ifx\csname urlprefix\endcsname\relax\def\urlprefix{URL }\fi
\providecommand{\eprint}[2][]{\url{#2}}

\bibitem{vonneumann}
von Neumann J 1932 {\em Mathematical Foundations of Quantum Mechanics\/} (New
  Jersey: Princeton University Press)

\bibitem{cohen-tannoudji}
Cohen-Tannoudji C, Diu B and Lalo\"e F 1977 {\em Quantum Mechanics\/} (New
  York: Wiley)

\bibitem{bohm}
Bohm D 1979 {\em Quantum Theory\/} (New York: Dover)

\bibitem{styer02}
Styer D 2002 {\em Am. J. Phys\/} {\bf 70} 288

\bibitem{bohm52a}
Bohm D 1952 {\em Phys. Rev.\/} {\bf 85} 166--179

\bibitem{holland:1993bk}
Holland P~R 1993 {\em The Quantum Theory of Motion: An Account of the De
  Broglie-Bohm Causal Interpretation of Quantum Mechanics\/} (Cambridge,
  England: Cambridge University Press)

\bibitem{einstein35}
Einstein A, Podolsky B and Rosen N 1935 {\em Phys. Rev.\/} {\bf 47} 777--780

\bibitem{ballentine70}
Ballentine L~E 1970 {\em Rev. Mod. Phys.\/} {\bf 42} 358--381

\bibitem{home92}
Home D and Whittaker M~A~B 1992 {\em Phys. Rep.\/} {\bf 210} 223--317

\bibitem{everett:1957pr}
Everett III H 1957 {\em Rev. Mod. Phys.\/} {\bf 29} 454

\bibitem{wheeler}
Wheeler J~A and Zurek W~H (eds) 1983 {\em Quantum Theory of Measurement\/}
  (Princeton: Princeton University Press)

\bibitem{bouda:2003ijmpa}
Bouda A 2003 {\em Int. J. Mod. Phys. A\/} {\bf 18} 3347

\bibitem{holland:2005ap}
Holland P 2005 {\em Ann. Phys.\/} {\bf 315} 505

\bibitem{Poirier:2010zza}
Poirier B 2010 {\em Chem. Phys.\/} {\bf 370} 4--14

\bibitem{holland10}
Holland P 2010 {\em Quantum Trajectories\/} ed Chattaraj P (Boca Raton: Taylor
  and Francis/CRC Press) chap~5, pp 73--85

\bibitem{poirier11nowaveCCP6}
Poirier B 2011 {\em Quantum Trajectories\/} ed Hughes K~H and Parlant G
  (Daresbury Laboratory: CCP6) p~6

\bibitem{Schiff:2012jcp}
Schiff J and Poirier B 2012 {\em J. Chem. Phys.\/} {\bf 136} 031102

\bibitem{poirier12ODE}
Parlant G, Ou Y~C, Park K and Poirier B 2012 {\em Comput. Theoret. Chem.\/}
  {\bf 990} 3

\bibitem{wiseman14prx}
Hall M~J~W, Deckert D~A and Wiseman H~M 2014 {\em Phys. Rev. X\/} {\bf 4}
  041013

\bibitem{poirier14prx}
Poirier B 2014 {\em Phys. Rev. X\/} {\bf 4} 040002

\bibitem{Poirier:2012gz}
Poirier B 2012 Trajectory-based theory of relativistic quantum particles
  (\href{http://arxiv.org/abs/1208.6260}{\textit{Preprint} arXiv:1208.6260
  [quant-ph]})

\bibitem{Ryder:1996bk}
Ryder L~H 1996 {\em Quantum Field Theory\/} 2nd ed (Cambridge, England:
  Cambridge University Press)

\bibitem{Wachter:2011bk}
Wachter A 2011 {\em Relativistic Quantum Mechanics\/} (Dordrecht: Springer)

\bibitem{thaller}
Thaller B 2005 {\em Advanced Visual Quantum Mechanics\/} (New York: Springer)

\bibitem{park12}
Park S~T 2012 {\em Phys. Rev. A\/} {\bf 86} 062105

\bibitem{Schweber:1961bk}
Schweber S~S 1961 {\em An Introduction to Relativistic Quantum Field Theory\/}
  (New York: Harper \& Row)

\bibitem{Gross:1999bk}
Gross F 1999 {\em Relativistic Quantum Mechanics and Field Theory\/} (New York:
  Wiley)

\bibitem{wyatt}
Wyatt R~E 2005 {\em Quantum Dynamics with Trajectories: Introduction to Quantum
  Hydrodynamics\/} (New York: Springer)

\end{thebibliography}
\bibliographystyle{iopart-num}

\end{document}